\documentclass[10pt,twocolumn,amsmath,amssymb,aps,prx,superscriptaddress,citeautoscript,longbibliography]{revtex4-2}

\usepackage{graphicx}% Include figure files
\usepackage{dcolumn}% Align table columns on decimal point
\usepackage{bm}% bold math
\usepackage{xcolor}
\usepackage{soul}
\usepackage{dcolumn}% Align table columns on decimal point
\usepackage{bm}% bold math
\usepackage{mathrsfs}
\usepackage{amsmath, amssymb}
\def\lesssim{\ \raise.3ex\hbox{$<$}\kern-0.8em\lower.7ex\hbox{$\sim$}\ }
\def\gesim{\ \raise.3ex\hbox{$>$}\kern-0.8em\lower.7ex\hbox{$\sim$}\ }

\def\up{\uparrow}
\def\down{\downarrow}

\begin{document}
\title{Non-Hermitian Renormalization Group from a Few-Body Perspective}
\author{Hiroyuki Tajima}
\affiliation{Department of Physics, Graduate School of Science, The University of Tokyo, Tokyo 113-0033, Japan}
\affiliation{RIKEN Nishina Center, Wako 351-0198, Japan}
\affiliation{Quark Nuclear Science Institute, The University of Tokyo, Tokyo 113-0033, Japan}

\author{Masaya Nakagawa}
\affiliation{Department of Physics, Graduate School of Science, The University of Tokyo, Tokyo 113-0033, Japan}

\author{Haozhao Liang}
\affiliation{Department of Physics, Graduate School of Science, The University of Tokyo, Tokyo 113-0033, Japan}
\affiliation{
    RIKEN Center for Interdisciplinary Theoretical and Mathematical Sciences (iTHEMS),
    Wako 351-0198, Japan}
\affiliation{Quark Nuclear Science Institute, The University of Tokyo, Tokyo 113-0033, Japan}

\author{Masahito Ueda}
\affiliation{Department of Physics, Graduate School of Science, The University of Tokyo, Tokyo 113-0033, Japan}
\affiliation{Institute for Physics of Intelligence, The University of Tokyo, 7-3-1 Hongo, Bunkyo-ku, Tokyo 113-0033, Japan}
\affiliation{RIKEN Center for Emergent Matter Science (CEMS), Wako, Saitama 351-0198, Japan}

\date{\today}
\begin{abstract}
Non-Hermiticity plays a fundamental role in open quantum systems and describes a wide variety of effects
of interactions with environments, including quantum measurement.
However, understanding its consequences in strongly interacting systems is still elusive due to the interplay between non-perturbative strong correlations and non-Hermiticity.
While the Wilsonian renormalization group (RG) method has been applied to tackle this problem, its foundation, based on the existence of the effective action and the partition function, is ill-defined.
In this paper, we establish a microscopic foundation of the RG method in non-Hermitian systems from a few-body perspective.
We show that the invariance of the scattering amplitude under RG transformations enables us to rigorously derive the non-Hermitian RG equation, 
giving a physically transparent interpretation of RG flows.
We discuss a detailed structure of such RG flows in a non-relativistic two-body system with an inelastic two-body loss, and show, in particular, its relation to a non-Hermitian quantum scale anomaly in two dimensions.
Our analysis suggests that non-Hermitian complex potentials often used in high-energy physics can be interpreted as being caused by quantum measurement, where the detection of elastically scattered particles updates the observer's knowledge, resulting in a nonunitary state change of the system. 
We apply our formalism to nuclear physics, find the emergence of a critical semicircle, which is a non-Hermitian generalization of a unitary limit, and show that several nuclei are located near the critical semicircle in the coherent neutron-nucleus scattering.
We also propose that the localized dineutron in two-neutron halo nuclei can be interpreted as a consequence of the quantum measurement effect on the imaginary potential associated with absorption into the core nucleus.
Our result bridges different contexts of non-Hermitian systems in high-energy and atomic, molecular, and optical physics, opening an interdisciplinary playground of non-Hermitian few-body physics.
\end{abstract}
%\begin{document}
\maketitle

\section{Introduction}
Non-Hermitian physics has attracted growing interest in diverse subfields of physics as an analytical continuation of conventional Hermitian quantum physics~\cite{ashida2020non}.
The birth of non-Hermitian physics dates back to Gamow’s description of nuclear alpha decay~\cite{Gamow1928}.
Since then, a complex-energy state called a resonant state has been studied extensively in nuclear physics~\cite{PhysRev.49.519,PhysRev.51.450,PhysRev.56.750}.
To describe the observed neutron-nucleus scattering cross section,
a non-Hermitian Hamiltonian called an optical model~\cite{PhysRev.75.1352,PhysRev.86.431} which involves a complex-valued potential 
was proposed.
After the development of the Feshbach projection formalism~\cite{feshbach1958unified,feshbach1962unified}, the construction of an optical potential from a microscopic nuclear force is among the most important issues in modern nuclear reaction theory~\cite{VARNER199157}. 

Meanwhile, non-Hermitian physics has made dramatic progress in the field of atomic, molecular, and optical (AMO) physics.
In the context of quantum measurement theory, the quantum trajectory description was discussed for continuous measurement~\cite{ueda1989probability,PhysRevA.41.3891,PhysRevLett.68.580,PhysRevA.45.4879,Daley04032014}, where the time evolution of the density matrix is governed by the nonunitary time evolution with a non-Hermitian Hamiltonian and quantum jump processes.
Remarkably, even without any jump process (e.g., no-count process of the photon detection in Refs.~\cite{ueda1989probability,PhysRevA.41.3891}), the time evolution of the monitored system is drastically different from that of an isolated quantum system due to an update of the observer's knowledge, namely, due to the fact that no jump occurs during the measurement.
In this regard, the nonunitary time evolution resulting from the measurement backaction can be regarded as a Bayesian inference of quantum states.
It should be noted that if the system possesses $\mathcal{PT}$-symmetry,
the eigenspectrum can be real for a $\mathcal{PT}$-unbroken phase~\cite{PhysRevLett.80.5243}.
This discovery leads to an important insight into our fundamental understanding of quantum mechanics, where a Hamiltonian having real-valued eigenenergies is not necessarily Hermitian.
Since then, various new concepts that are absent in Hermitian systems, such as non-Hermitian topology and exceptional points, have been discussed~\cite{PhysRevX.8.031079,PhysRevX.9.041015}.

Despite these advances,
strongly interacting non-Hermitian systems 
present a major theoretical challenge.
In this context, the renormalization group (RG) method is a powerful tool to incorporate strong-coupling effects in a non-perturbative manner.
The conceptual idea of RG was initiated in quantum electrodynamics~\cite{PhysRev.95.1300} and generalized to arbitrary renormalizable theories~\cite{PhysRevD.2.1541,symanzik1970small}, leading to the discovery of asymptotic freedom in quantum chromodynamics~\cite{PhysRevLett.30.1343,PhysRevLett.30.1346}.
In condensed matter physics, the block-spin RG approach developed by Kadanoff gives a clear physical picture of the RG flow in the Ising model~\cite{PhysicsPhysiqueFizika.2.263}.
Later, Wilson established the modern RG method by studying how the effective action changes as one lowers the ultraviolet cutoff~\cite{RevModPhys.55.583}.
In this regard, the RG approach was also applied to reveal 
unique features of strongly interacting non-Hermitian systems such as the breakdown of the $c$ theorem in conformal field theory (CFT)~\cite{fendley1993massless,fendley1993massless2}, parity-time symmetric quantum critical phenomena~\cite{ashida2017parity}, 
phase transitions in complex CFT~\cite{gorbenko2018walking,PhysRevLett.124.051602,yang2026asymptotic,yamamoto2026complex},
and the semicircle distribution of Yang-Lee zeros in non-Hermitian superconductivity~\cite{PhysRevLett.131.216001}.
However, since the Wilsonian perturbative RG approach presupposes the existence of the partition function~\cite{RevModPhys.55.583}, its foundational ground is elusive in non-Hermitian systems since the equilibrium Gibbs state is ill-defined. 

In this paper, we establish the microscopic foundation of the RG approach in non-Hermitian few-body systems, thereby bridging non-Hermitian physics in AMO and nuclear systems where strong interparticle interactions play a fundamental role.
Regarding the RG approach to the nuclear force, an effective low-momentum interaction $V_{{\rm low}\, k}$~\cite{bogner2003model}, which is obtained by the phase shift equivalence during the RG evolution, is a useful tool to study the low-energy universality of nuclear forces.
Moreover, $V_{{\rm low}\, k}$ does not rely on the existence of the effective action and hence the partition function.
In this sense, although the RG scale corresponds to the ultraviolet cutoff, the idea of this RG approach is similar to the Callan-Symanzik formalism~\cite{PhysRevD.2.1541,symanzik1970small} which does not require the existence of the equilibrium Gibbs state.

Another advantage of non-Hermitian few-body problems is that the postselection can be performed in cold atom experiments with tunable interactions. 
For instance, a non-Hermitian few-body system can be realized by collecting the data of survival states under particle loss via recent techniques such as matter-wave magnification~\cite{wdjr-m2hg} and by absorption spectroscopy~\cite{PhysRevLett.129.093001} under photoassociation two-body loss~\cite{RevModPhys.78.483}. 
By comparing the RG flow and the exact solution in non-Hermitian few-body systems,
we can provide a rigorous physical interpretation of the non-Hermitian RG flow.

\begin{figure*}[t]
    \centering
    \includegraphics[width=0.9\linewidth]{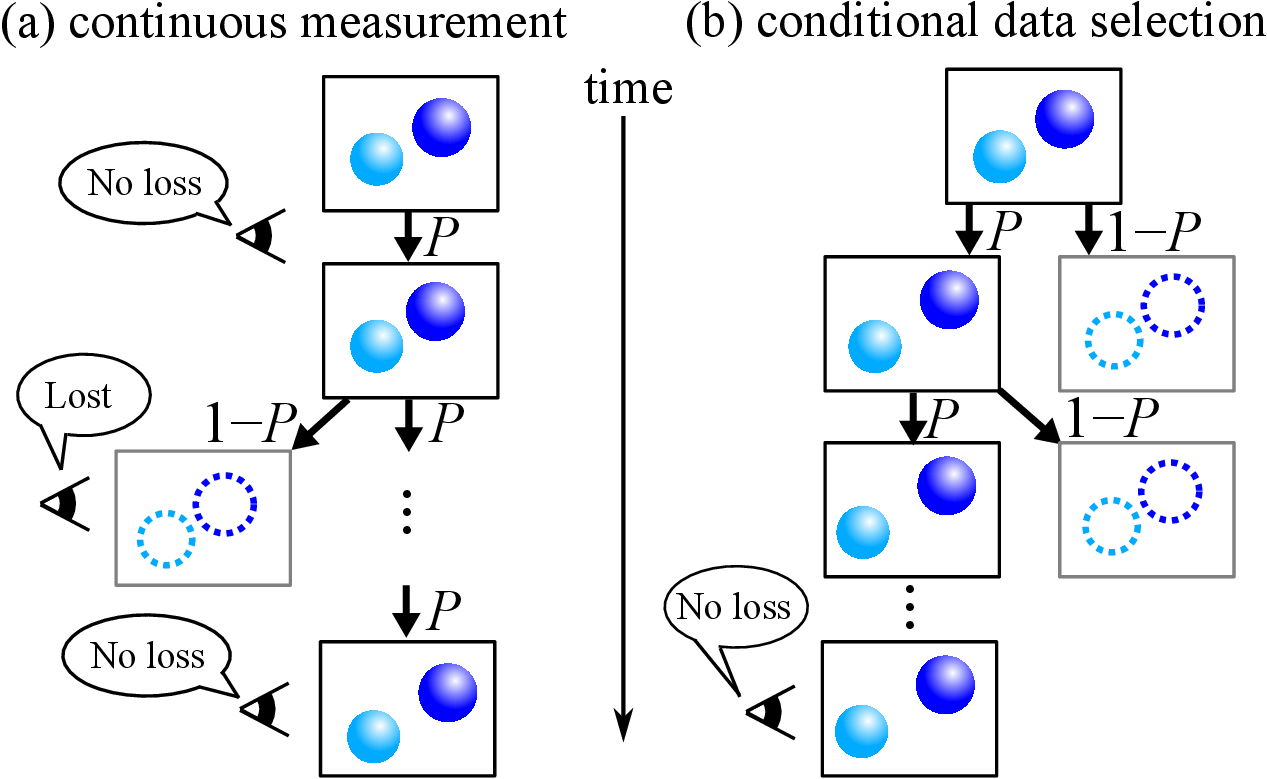}
    \caption{Schematic illustrations of (a) continuous measurement discussed in AMO physics and (b) conditional data selection in high-energy experiments,
    through projection $P$ on survival states with inelastic loss.
    In (a), a monitored system undergoes nonunitary time evolution due to continuous measurement.
    Even in the absence of the jump process, the state is affected by measurement backaction due to the projector $P$. In (b), after the scattering event, the measurement data are conditionally selected for a specific channel (i.e., elastic scattering channel) which has undergone no jump processes characterized by the projector $P$.
    }
    \label{fig:1}
\end{figure*}

Against these backdrops, we study a non-Hermitian few-body system with inelastic two-body collisions.
The purpose of this paper is twofold:
\begin{enumerate}
    \item To establish a microscopic foundation of the non-Hermitian RG approach on the basis of the RG scale invariance of the scattering amplitude in a few-body system. 
    \item To explore quantum-measurement effects in nuclear physics by shedding light on a similarity between postselected AMO and nuclear experiments.
\end{enumerate}
We develop the RG framework that relies on the invariance of the scattering amplitude, rather than the equilibrium partition function, in contrast to the Wilsonian RG approach, which is based on the perturbation upon the equilibrium Gibbs state at each RG scale.
By doing so, we establish the foundation of the non-Hermitian RG approach widely used in recent studies of strongly coupled non-Hermitian AMO systems.
Importantly, the non-Hermitian few-body problem is relevant to nuclear experiments.
While measurement protocols are different in AMO and high-energy experiments, 
there exists a notable similarity in the non-Hermitian description of quantum states as shown in Fig.~\ref{fig:1}.
In this context, however, quantum-measurement perspectives have not been taken into account in nuclear physics.
For instance, the primary purpose of the optical model is to describe the observed scattering cross section in terms of a mean-field optical potential.
Little attention has been paid to the concomitant nonunitary state change of the system.
Also, the quantum-measurement effect can be a crucial key to understanding the observation of multi-neutron states in halo nuclei~\cite{zhukov1993bound,tanihata1996neutron,tanihata2013recent}, because the multi-neutron subsystem can be regarded as an open system subject to the formation of a bound state with a core nucleus.
Representative examples would be recent observations of a tetraneutron state,
where two experiments report its resonance~\cite{PhysRevLett.116.052501,duer2022observation} and another experiment indicates a bound tetraneutron~\cite{faestermann2022indications}.
These observations are in contrast to theoretical calculations showing the absence of a localized tetraneutron state~\cite{marques2021quest}.
The different experimental setups among them can make a difference on the results due to the quantum-measurement effect.

To be specific, we apply our framework to non-relativistic two-body problems in different spatial dimensions to discuss non-Hermitian RG flows and quantum measurement effects.
It is known that a zero-range attractive interaction, which classically possesses scale invariance in two dimensions, induces a quantum scale anomaly~\cite{PhysRevD.5.377} as an emergence of a new energy scale associated with a two-body bound state~\cite{PhysRevLett.108.185303}.
Also, the two-body contact interaction in two dimensions is among the rare examples of a nonrelativistic system that displays asymptotic freedom~\cite{PhysRevLett.93.250408}.
In the context of cold atoms, the quantum scale anomaly has already been observed via the measurement of a breathing mode~\cite{PhysRevA.55.R853,PhysRevLett.105.095302} in two-dimensional Hermitian many-body systems~\cite{PhysRevLett.121.120401,PhysRevLett.121.120402,PhysRevX.9.021035}.
However, its non-Hermitian counterpart has been elusive because of the difficulty in postselecting no-jump processes of the collective mode under dissipation.
Nevertheless, the non-Hermitian RG approach enables us to investigate the fate of the quantum scale anomaly in non-Hermitian two-body systems.
Since the two-body binding energy takes the non-perturbative form with respect to the running two-body coupling exhibiting the asymptotic freedom, it is a good testing ground for the non-Hermitian RG approach to the non-perturbative strong-coupling effect.

Since non-Hermitian two-body interaction is directly related to the optical potential, our study can pave the way to explore the quantum measurement effect in high-energy experiments such as nuclear reactions.
In this regard, we examine the three-dimensional two-body case relevant to the coherent neutron-nucleus scattering~\cite{sears1992neutron,squires1996introduction} where the incident neutron can be absorbed into the nucleus, leading to inelastic two-body loss.
The relationship between the non-Hermitian RG flow and the neutron-nucleus scattering will be discussed in Sec.~\ref{sec:6}.

The non-Hermitian few-body description is useful for understanding multineutron subsystems in neutron-rich halo nuclei~\cite{zhukov1993bound,tanihata1996neutron,tanihata2013recent}.
So far, 
a variety of Borromean two-neutron halo nuclei such as $^6$He~\cite{ter1998two},
$^{11}$Li~\cite{PhysRevLett.96.252502}, $^{19}$B~\cite{PhysRevLett.124.212503},
$^{22}$C~\cite{PhysRevLett.104.062701}, and
$^{29}$F~\cite{PhysRevLett.124.222504}
have been observed, where ``Borromean" indicates that a three-body bound state exists 
in a region where no two-body bound states can exist. 
The three-body analysis (i.e., two neutrons and core nucleus) suggests localized dineutron correlations in contrast to an isolated two-neutron system which is unbound~\cite{PhysRevC.72.044321,PhysRevC.80.031301,PhysRevLett.99.022506}.
We show that the localization of the two-neutron subsystem in halo nuclei can be understood as a quantum measurement effect; i.e., nonunitary state change due to two-neutron non-absorption to the core nucleus
when the probability of two-neutron absorption is nonzero.

This paper is organized as follows.
In Sec.~\ref{sec:2}, we provide a general formalism of non-Hermitian few-body systems and show the role of Bayesian inference of an observer in the nonunitary time evolution.
In Sec.~\ref{sec:3}, we present the non-Hermitian RG approach derived from the invariance of the scattering amplitude (i.e., scattering $T$ matrix) under the RG scale transformation.
We apply the non-Hermitian RG to a two-body system with inelastic two-body loss, which
can be engineered by photoassociation~\cite{tomita2017observation,PhysRevLett.132.263401} in cold atoms, and compare the RG flows in different spatial dimensions.
To understand the non-Hermitian RG flow,
in Sec.~\ref{sec:4}, we discuss its relation to the resonant state obtained by solving the non-Hermitian two-body Schr\"{o}dinger equation.
In Sec.~\ref{sec:5}, we present non-Hermitian quantum scale anomaly as a representative example that can be understood from non-trivial RG flow.
In Sec.~\ref{sec:6}, we examine the neutron-nucleus scasttering in three spatial dimensions.
In Sec.~\ref{sec:7}, we discuss the relationship between the quantum-measurement effect and dineutron correlations in two-neutron Borromean nuclei.
Finally, we give a summary in Sec.~\ref{sec:8}.
Throughout the paper, we set $\hbar=1$.

\section{nonunitary time evolution of few-body systems}
\label{sec:2}
We consider a scattering problem where two particles undergo elastic or inelastic collisions, by the latter of which the scattered particles are lost from the system.
Such an inelastic scattering is often observed in cold-atom experiments, and it is also described by the optical potential in nuclear physics~\cite{VARNER199157}.
Here, we describe the time evolution of the density matrix $\rho(t)$ of the system under inelastic collisions by using the Gorini-Kossakowski-Sudarshan-Lindblad equation~\cite{gorini1976completely,lindblad1976generators}
\begin{align}
    \frac{d\rho(t)}{dt}&=-i[H,\rho(t)]\cr
    &\, \,
    -g_{i}\int d\bm{r}\left[\left\{L_{\bm{r}}^\dag L_{\bm{r}},\rho(t)\right\}-2L_{\bm{r}}\rho(t)L_{\bm{r}}^\dag\right],
\end{align}
where $H$ is the Hermitian Hamiltonian of the system of our interest.
The inelastic loss is described by
the Lindblad operator $L_{\bm{r}}$ with a strength parameter $g_{i}\in\mathbb{R}$ ($g_{i}>0$).
Considering the scattering problem in the presence of an inelastic collision, one can decompose the density matrix as
\begin{align}
    \rho(t)=\rho_{\rm s}(t)+\rho_0(t),
\end{align}
where $\rho_{\rm s}(t)$ is an unnormalized density matrix of the subspace where the inelastic loss does not occur, and $\rho_0(t)$ is the density matrix of the subspace undergoing inelastic loss.
For instance, in a two-body problem, $\rho_{\rm s}(t)$ and $\rho_0(t)$ correspond to the unnormalized density matrices of two- and zero-particle states, respectively.
Using a projector $P$ 
onto the two-particle subspace,
we get 
\begin{align}
    \rho_{\rm s}(t)=P\rho(t)P,
\end{align}
\begin{align}
    \rho_0(t)=(1-P)\rho(t)(1-P).
\end{align}
The fact that $L_{\bm{r}}$ acts on only $\rho_{\rm s}(t)$ with a reduction of the particle number leads to
\begin{align}
\label{eq:5}
    \frac{d\rho_{\rm s}(t)}{dt}=-i\left[H_{\rm eff}\rho_{\rm s}(t)-\rho_{\rm s}(t)H_{\rm eff}^{\dag}\right],
\end{align}
\begin{align}
    \frac{d\rho_0(t)}{dt}=2g_{i}\int d\bm{r}\,
    L_{\bm{r}}\rho_{\rm s}(t)L_{\bm{r}}^\dag,
\end{align}
where 
\begin{align}
\label{eq:7}
    H_{\rm eff}&=H-ig_{i}\int d\bm{r}\, L_{\bm{r}}^\dag L_{\bm{r}}%\cr
  % &=\int d\bm{r}\,\psi_{\sigma}^\dag(\bm{r})\left(-\frac{\nabla^2}{2m}\right)
  %  \psi_{\sigma}(\bm{r})
  %  \cr
  %  &\quad +g_{\Lambda}
  %  \int d\bm{r}\,\psi_{\uparrow}^\dag(\bm{r})\psi_{\downarrow}^\dag(\bm{r})\psi_{\downarrow}(\bm{r})\psi_{\uparrow}(\bm{r}) 
\end{align}
is a non-Hermitian effective Hamiltonian describing the nonunitary time evolution.
%with a complex-valued coupling constant $g_{\Lambda}=g_{r,\Lambda}-ig_{i,\Lambda}$.

After the scattering event, we collect the two particles by a detector and measure the scattering cross section. We assume that particles after a scattering event can be detected with certainty if they are not lost from the system, and that we cannot detect them if they are lost due to an inelastic collision.
This situation corresponds to an idealized detector with perfect efficiency.
Suppose that we
perform a measurement at a certain time and detect particles, which means that they have not been lost during
the scattering process. 
Then, the density matrix of the subsystem that is probed by the detector is given by 
\begin{align}
\label{eq:8}
\tilde{\rho}_{\rm s}(t)=\frac{P\rho(t)P}{{\rm Tr}\left[P\rho(t)P\right]}\equiv \frac{\rho_{\rm s}(t)}{{\rm Tr}[\rho_{\rm s}(t)]},    
\end{align}
which can be obtained from the nonunitary time evolution under $H_{\rm eff}$, governed by Eq.~\eqref{eq:5}. 
This is because the detection event tells us that the particles have not been lost from the system; the update of the observer's knowledge about the system leads to a nonunitary state change due to measurement backaction.

\section{Non-Hermitian renormalization group formalism}
\label{sec:3}
We now derive the non-Hermitian RG equation from the scattering amplitude.
Here, let us consider the simplest case of a two-body inelastic collision.
In particular, if the initial state of the system is given by a two-particle pure state $\rho(t\rightarrow-\infty) = |\Phi_0\rangle \langle\Phi_0|$, we can write the density matrix as $\rho_{\rm s}(t) = |\Psi(t)\rangle \langle\Psi(t)|$, where $|\Psi(t)\rangle$
obeys the non-Hermitian Schr\"{o}dinger equation
\begin{align}
    i\frac{d}{dt}|\Psi(t)\rangle=H_{\rm eff}|\Psi(t)\rangle.
\end{align}
Incidentally, the norm of $|\Psi(t)\rangle$ decreases in time due to the inelastic collision as
\begin{align}
    \frac{d}{dt}\langle\Psi(t)|\Psi(t)\rangle=i\langle\Psi(t)|(H_{\rm eff}^\dag-H_{\rm eff})|\Psi(t)\rangle<0.
\end{align}
In this regard, $\tilde{\rho}_{\rm s}(t)$ is normalized by $\langle\Psi(t)|\Psi(t)\rangle\equiv {\rm Tr}[\rho_{\rm s}(t)]$ as shown in Eq.~\eqref{eq:8}.

As in the conventional scattering theory, it is useful to write $H_{\rm eff}=H_0+V$, where
$H_0$ is the kinetic term
and
$V$ is the non-Hermitian two-body interaction involving both elastic and inelastic processes.
We then obtain the scattering amplitude by solving the Lippmann-Schwinger equation of $|\Psi(t)\rangle=e^{-iEt}|\Psi_0\rangle$
as (see also Appendix~\ref{app:a})
\begin{align}
\label{eq:10}
  |\Psi_0\rangle=|\Phi_0\rangle +\frac{1}{E+i\delta-H_0}V|\Psi_0\rangle, 
\end{align}
with a real-valued eigenenergy $E\in\mathbb{R}$ (i.e., $H_0|\Phi_0\rangle=E|\Phi_0\rangle$), where
$\delta$ in Eq.~\eqref{eq:10} is an infinitesimally small positive number.
The scattering amplitude is directly related to the scattering $T$ matrix given by~\cite{PhysRev.51.450}
\begin{align}
\label{eq:11}
T(E)=V+V\frac{1}{E+i\delta-H_0}T(E).    
\end{align}
Note that Eq.~\eqref{eq:11} holds regardless of whether $V$ is Hermitian or non-Hermitian.
From the asymptotic form of the wave function,
we obtain the scattering amplitude~\cite{newton2013scattering}
\begin{align}
\label{eq:13}
    f_k(\theta)=-\frac{\mu k^{\frac{d-3}{2}}}{(2\pi)^{\frac{d-1}{2}}}
    \langle\bm{k}'|T(E)|\bm{k}\rangle,
\end{align}
where $d$ is a spatial dimension,
$k=\sqrt{2\mu E}$ with a reduced mass $\mu$, and
$\bm{k}$ and $\bm{k}'$ are the incoming and outgoing relative momenta with a relative angle $\theta$,
satisfying $|\bm{k}|=|\bm{k}'|= k$.

It should be noted that Eq.~\eqref{eq:13} is valid even for the non-Hermitian interaction because $f_k(\theta)$ represents the amplitude of the outgoing wave function at a large distance \textit{without} the loss process.
To understand this, it is convenient to see the probability current $\bm{j}(\bm{r})$ of the scattering wave function $\varphi(\bm{r})=e^{ikz} +f_k(\theta)e^{ikr}/r$ at $d=3$
where we take the direction of the incoming wave as the $z$ axis without loss of generality.
We obtain
\begin{align}
    \bm{j}(\bm{r})&=\frac{1}{2\mu}{\rm Im}\left[\varphi^*(\bm{r})\nabla\varphi(\bm{r})\right]\cr
    &=\frac{k}{2\mu}\bm{e}_z+\frac{k}{2\mu}\frac{|f_{k}(\theta)|^2}{r^2}\bm{e}_r
    +\frac{k}{2\mu r}\cr
    &\,\times
    {\rm Re}[f_k(\theta)e^{ik(r-z)}\bm{e}_r+f_k^*(\theta)e^{-ik(r-z)}\bm{e}_z],
\end{align}
where $\bm{e}_{r}$ and $\bm{e}_z$ are the unit vectors parallel to $\bm{r}$ and the $z$ axis, respectively.
The probability flux $\Phi(R)$ at a large distance $R$ from the scattering point is obtained by the integration of $\bm{j}(\bm{r})$ on the surface $S(R)$ with radius $R$ as 
\begin{align}
\label{eq:phi}
    \Phi(R)&=\int_{S(R)}R^2d\Omega\,\bm{e}_r\cdot\bm{j}(\bm{r})\cr
    &=\frac{k}{2\mu}\sigma_{\rm el}-\frac{2\pi}{\mu}{\rm Im}\left[f_k(0)\right],
\end{align}
where $\sigma_{\rm el}=\int_{S(\Omega)}d\Omega\,|f_k(\theta)|^2$
is the elastic scattering cross section, that is, the amplitude obtained by observing all the outgoing wave functions.
The second term of Eq.~\eqref{eq:phi} represents the interference between the incoming and scattering waves at the forward scattering ($\theta=0$) in the limit of $R\rightarrow \infty$.
If there is no inelastic loss, we should get $\Phi(R)=0$, indicating the optical theorem $\sigma_{\rm el}=\frac{4\pi}{k}{\rm Im}[f_k(0)]$ as a conservation of the probability.
However, in the non-Hermitian case, this relation is broken as $\Phi(R)< 0$ because the probability current flows into the inside of $S(R)$ due to the inelastic loss at the scattering region.
Accordingly, one may define the absorption (inelastic) cross section $\sigma_{\rm abs}$ and the total one $\sigma_{\rm tot}$ as
\begin{align}
    \sigma_{\rm abs}=-\frac{2\mu}{k}\Phi(R)\equiv\frac{4\pi}{k}{\rm Im}[f_k(0)]-\sigma_{\rm el},
\end{align}
and
\begin{align}
\label{eq:sigma_tot}
    \sigma_{\rm tot}=\sigma_{\rm el}+\sigma_{\rm abs}\equiv \frac{4\pi}{k}{\rm Im}[f_k(0)].
\end{align}
We thus obtain an unambiguous interpretation of the scattering amplitude and the cross sections even in non-Hermitian systems.

Consequently, the non-Hermitian RG equation can be obtained as the condition of the invariance of
the scattering amplitude under the scale transformation.
We emphasize that this is in sharp contrast with the Wilsonian RG method based on the effective action and the many-body partition function whose definition is not straightforward in non-Hermitian systems.
For a given RG scale $\Lambda$, we have $\frac{\partial}{\partial\Lambda}f_k(\theta)=0$, leading to
\begin{align}
\label{eq:12}
    \frac{\partial}{\partial \Lambda}T(E)=0,
\end{align}
with a fixed $E$,
which is equivalent to the Callan-Symanzik equation~\cite{PhysRevD.2.1541,symanzik1970small} due to the fact that the $T$ matrix is also obtained from the amputated two-body Green's function~\cite{kopietz2010introduction}.
Introducing the complex-valued running coupling constant $g_\Lambda=g_{r,\Lambda}-ig_{i,\Lambda}$, we get its RG flow equation from Eq.~\eqref{eq:12}.
It should be noted that Eq.~\eqref{eq:12} is often used in the construction of the low-momentum effective interaction $V_{{\rm low}\, k}$ via the RG transformation~\cite{bogner2003model}.

While we mainly consider two-body subsystems in an open environment, Eq.~\eqref{eq:12} can be applied to
general cases with more than two particles. Indeed, the celebrated RG limit cycle in Efimov physics can be obtained from the RG transformation of the three-body (atom-dimer) scattering $T$ matrix~\cite{PhysRevLett.82.463}.
Even in many-body environments, a few-body subsystem plays a crucial role in understanding strong correlations within the total system, as it is evident that our approach with a two-body subsystem can be applied to Borromean two-neutron halo nuclei in Sec.~\ref{sec:7}.
In conventional many-body theories, we exploit single- and few-particle Green's functions~\cite{dickhoff2004self,dickhoff2008many}.
From an open-system perspective, these few-particle Green's functions describe the dynamics of a few-body subsystem.
In this context, the $T$ matrix is obtained from the amputated few-particle Green's function~\cite{kopietz2010introduction}.
Although we need to use additional approximations on the background medium to obtain the $T$ matrix in many-body environments, in principle, this approach does not require the Gibbs state, in contrast to the Wilsonian RG method. (To be specific, the expectation value defining the Green’s functions is not necessarily taken in a Gibbs state~\cite{mahan2013many,rammer2007quantum}.)
Our non-Hermitian RG method thus has a clear physical meaning
and extendable to more complicated many-body systems under dissipation.

Note that the present approach is similar to the poor man's scaling method applied to the non-Hermitian Kondo model~\cite{PhysRevLett.121.203001,PhysRevB.98.085126,PhysRevB.107.235153,PhysRevB.111.045124}.
While the RG equation in the non-Hermitian Kondo model is obtained approximately because of the background Fermi sea, the RG equation in this paper is exact at the two-body level, and therefore, by comparing it with the exact two-body calculation, we can gain physical insights into the non-Hermitian RG flow.

\section{Demonstration of non-Hermitian renormalization-group flow in a non-relativistic two-body problem}
\label{sec:4}
To be specific, in what follows, let us consider a non-relativistic two-component fermionic system described 
by a Hermitian Hamiltonian
\begin{align}
\label{eq:14}
    H&=
    \int d\bm{r}\,\sum_{\sigma=\up,\down}\psi_{\sigma}^\dag(\bm{r})\left(-\frac{\nabla^2}{2m_\sigma}\right)
    \psi_{\sigma}(\bm{r})
    \cr
    &\quad +g_{r,\Lambda}
    \int d\bm{r}\,\psi_{\uparrow}^\dag(\bm{r})\psi_{\downarrow}^\dag(\bm{r})\psi_{\downarrow}(\bm{r})\psi_{\uparrow}(\bm{r}), 
\end{align}
where $\psi_{\sigma}(\bm{r})$ is the field operator of a fermion with mass $m_\sigma$ and internal state $\sigma=\up,\down$ (e.g., spin).
For the nucleon-nucleus scattering case, one may assume that $\sigma=\up$ and $\sigma=\down$ represent a nucleon and a nucleus, respectively.
We note that for two distinguishable particles, their statistics are not important.
The last term in Eq.~\eqref{eq:14} describes the two-body interaction with the real-valued coupling $g_{r,\Lambda}\in\mathbb{R}$ at the RG momentum scale $\Lambda$.
Furthermore, we consider the inelastic two-body loss described by $L_{\bm{r}}=\psi_{\downarrow}(\bm{r})\psi_{\uparrow}(\bm{r})$.
Taking the running imaginary coupling as $g_i\rightarrow g_{i,\Lambda}$ in Eq.~\eqref{eq:7}, we obtain the non-Hermitian effective Hamiltonian
\begin{align}
\label{eq:15}
    H_{\rm eff}&=
    \int d\bm{r}\,\sum_{\sigma=\up,\down}\psi_{\sigma}^\dag(\bm{r})\left(-\frac{\nabla^2}{2m_\sigma}\right)
    \psi_{\sigma}(\bm{r})
    \cr
    &\quad +g_{\Lambda}
    \int d\bm{r}\,\psi_{\uparrow}^\dag(\bm{r})\psi_{\downarrow}^\dag(\bm{r})\psi_{\downarrow}(\bm{r})\psi_{\uparrow}(\bm{r}), 
\end{align}
with a complex-valued running coupling $g_\Lambda=g_{r,\Lambda}-ig_{i,\Lambda}$.

Following the procedure shown in Sec.~\ref{sec:3}, one can obtain the two-body $T$ matrix
\begin{align}
\label{eq:16}
    T(E)=\frac{g_{\Lambda}}{1-g_{\Lambda}\Pi(E)},
\end{align}
where
\begin{align}
\label{eq:17}
    \Pi(E)&=\sum_{|\bm{p}|\leq \Lambda}\frac{1}{E+i\delta-p^2/2\mu}\cr
    &\equiv\frac{S_d}{(2\pi)^d}\int_0^{\Lambda}\frac{p^{d-1}dp}{E+i\delta-p^2/2\mu}
\end{align}
is the two-body propagator.
In Eq.~\eqref{eq:17}, the ultraviolet cutoff is set to $\Lambda$ and $\mu=m_{\up}m_{\down}/(m_{\up}+m_{\down})$ is the reduced mass.
In the second line of Eq.~\eqref{eq:17}, the momentum summation is replaced with the integration where  $S_d=2\pi^{d/2}/\Gamma(d/2)$ is the surface area of a $d+1$-dimensional unit sphere with $\Gamma(x)$ being the gamma function.
Performing the momentum integration, we obtain
the explicit form of $\Pi(E)$ as
\begin{align}
\label{eq:18}
    \Pi(E)
    &=\left\{
    \begin{array}{cc}
        -\frac{\mu}{\pi\sqrt{2\mu E}}\ln\left(\frac{\Lambda^2-\sqrt{2\mu E+i\delta}}{\Lambda^2+\sqrt{2\mu E}}\right) &  (d=1);\\
        -\frac{\mu}{2\pi}\ln\left(\frac{2\mu E+i\delta-\Lambda^2}{2\mu E+i\delta}\right) &  (d=2); \\
       -\frac{\mu\Lambda}{\pi^2}-i\frac{\mu}{2\pi}\sqrt{2\mu E}  &  (d=3).
    \end{array}
    \right. 
\end{align}
Combining Eqs.~\eqref{eq:12}, \eqref{eq:16}, and \eqref{eq:18}
under the assumption of $E\ll \Lambda^2/2\mu$, we obtain the non-Hermitian RG equation for a $d$-dimensional system as
\begin{align}
\label{eq:19}
    \frac{dg_{\Lambda}}{d\Lambda}&=
    \frac{2\mu S_d}{(2\pi)^d\Lambda^{3-d}}
    g_\Lambda^2.
\end{align}
Note that, while we consider the momentum cutoff as a RG scale,
the RG equation does not formally depend on the choice of the RG scale in the sense of the Callan-Symanzik equation~\cite{collins1984renormalization}.
We show the case with a different RG regularization in Appendix~\ref{app:b}.

We note that the scattering amplitude with complex-valued $s$-wave scattering length $a$, which is  relevant to nuclear experiments such as
nucleon-nucleus scattering~\cite{sears1992neutron},
$\eta$-nucleon scattering~\cite{RevModPhys.91.015003} and
kaonic atoms~\cite{RevModPhys.91.025006},
can be obtained within the present formalism.
Using Eq.~\eqref{eq:13} in $d=3$,
we obtain
\begin{align}
\label{eq:25}
    f_k(\theta)
    %&=
    %-\frac{\mu}{2\pi}\left(\frac{1}{g_\Lambda}+\frac{\mu\Lambda}{\pi^2}+i\frac{\mu k}{2\pi}\right)^{-1}
    %\cr
    &=\left(
    -\frac{2\pi}{\mu g_{\Lambda}}-\frac{2\Lambda}{\pi}-ik
    \right)^{-1}\cr
    &\equiv\left(-\frac{1}{a}-ik\right)^{-1}.
\end{align}
We thus obtain
\begin{align}
\label{eq:26}
    a=\left(\frac{2\pi}{\mu g_\Lambda} +\frac{2\Lambda}{\pi}\right)^{-1}\in\mathbb{C}.
\end{align}

To see the impact of the imaginary running coupling in the non-Hermitian RG equation, it is illuminating to consider the case with pure imaginary coupling $g_{\Lambda}=-ig_{i,\Lambda}$ ($g_{i,\Lambda}\in\mathbb{R}$ and $g_{i,\Lambda}>0$).
In such a case, the system evolves non-unitarily due to the imaginary coupling and could be drastically different from an isolated non-interacting two-particle system.
In particular, even for $g_{\Lambda}=-ig_{i,\Lambda}$,
the real part of the running coupling is generated in the RG flow. 
In other words, in contrast to an isolated system,
the two-particle interaction is generated by the measurement backaction.
In analogy with the nonunitary time evolution, where
the readout information that such a process did not occur leads to a nonunitary change of the density matrix in time, the emergent real-valued running coupling can be regarded as a consequence of the Bayesian inference in the RG flow.
In this sense,
we call
the terms induced by nonzero $g_{i,\Lambda}$ in Eq.~\eqref{eq:19} a \textit{Bayesian inference term}.
The resulting scattering amplitude can be nonzero based on Eq.~\eqref{eq:25}, indicating that the elastic channel exists as a consequence of the Bayesian inference term even though the original interaction does not have the elastic real part.

\begin{figure*}[tb]
    \centering
    \includegraphics[width=1\linewidth]{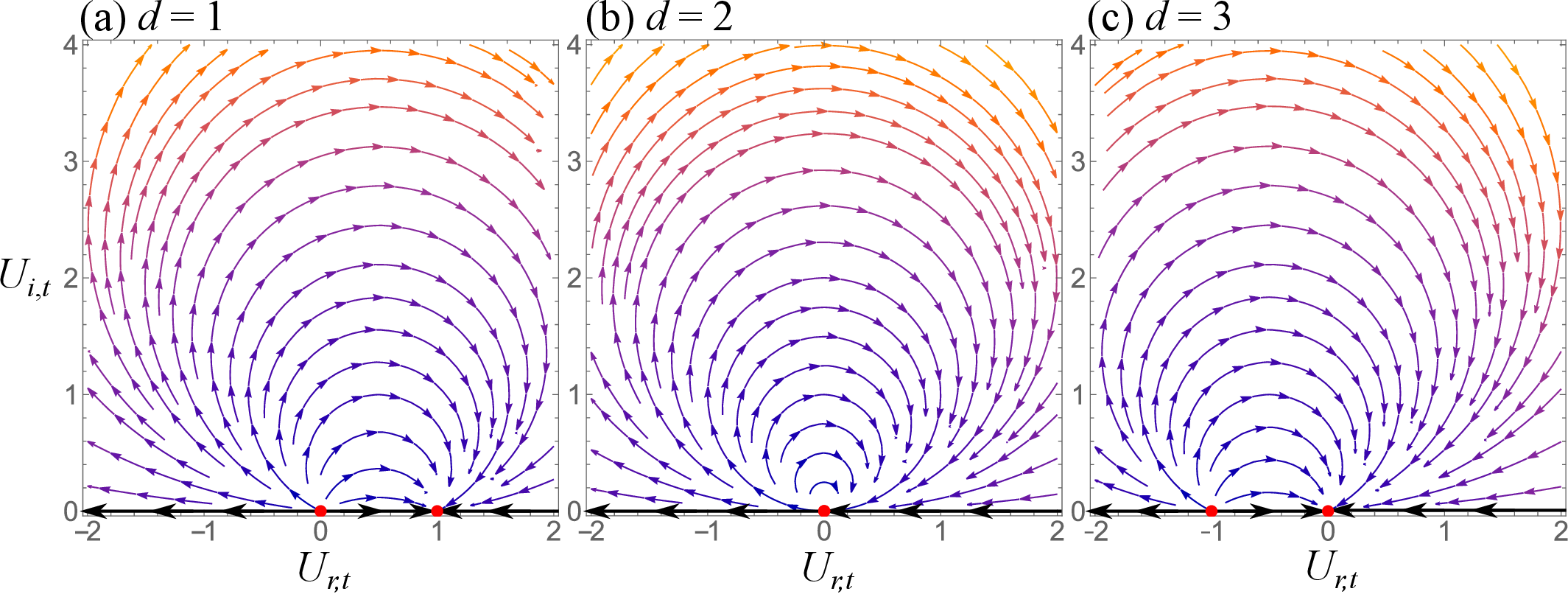}
    \caption{Non-Hermitian RG flow of the dimensionless complex-valued coupling $U_t=U_{r,t}-iU_{i,t}$ in (a) $d=1$, (b) $d=2$, and (c) $d=3$.
    The red points denote the fixed points.
    At $U_{i,t}=0$ (solid thick lines),
    the RG flow shows $U_{r,t\rightarrow 0}\rightarrow-\infty$ at $U_{r,t}< 0$ in $d=1,2$,  corresponding to the emergence of a two-body bound state. Note that the flow to $U_{r,t\rightarrow\infty}=-\infty$ is found for $U_{r,t}<-1$ in $d=3$, where the unstable fixed point $U_{r,t}=-1$ is the unitary limit~\cite{PhysRevA.75.033608}.
    }
    \label{fig:2}
\end{figure*}

It is useful to introduce the RG flow time $t$ as $\Lambda_{t}=\Lambda_0 e^{-t}$ where $\Lambda_0$ is the UV cutoff momentum.
The RG equation of the dimensionless running coupling 
\begin{align}
    U_{r/i,t}=\frac{2S_d}{(2\pi)^{d}}\mu g_{r/i,\Lambda}\Lambda^{d-2}
\end{align}
is given by
\begin{align}
\label{eq:4}
    \frac{dU_{r,t}}{dt}&=
    -U_{r,t}^2-(d-2)U_{r,t}+U_{i,t}^2,
    \cr
    \frac{dU_{i,t}}{dt}&=
    -(2U_{r,t}+d-2)U_{i,t},
\end{align}
indicating that the RG flow changes its direction in the complex plane of $U_t=U_{r,t}-iU_{i,t}$ due to the Bayesian inference term associated with $U_{i,t}$.
We emphasize that the present non-Hermitian RG equation is exact in this two-body problem.
Indeed, Eq.~\eqref{eq:4} is equivalent to the RG equation derived for a real-valued coupling with $U_{i,t=0}=0$
by means of the Wilsonian perturbative RG method~\cite{PhysRevB.46.11749,PhysRevA.75.033608}.

Figure~\ref{fig:2} shows the non-Hermitian RG flow in the plane of $U_{r,t}$ and $U_{i,t}$ in different spatial dimensions.
In particular, $d=2$ is the critical dimension where the quantum scale anomaly occurs in the Hermitian counterpart~\cite{PhysRevLett.108.185303}.
While $U_{r,t}\rightarrow-\infty$ is found at $t\rightarrow \infty$ in the case with $U_{i,t}=0$, corresponding to the two-body bound state, a loop structure emerges in the non-Hermitian RG flow.
We note here that a looped RG flow is also found in the non-Hermitian Kondo effect~\cite{PhysRevLett.121.203001}.
In the present case, in comparison with the cases of $d=1$ and $d=3$ that exhibit unidirectional complex-valued RG flows between two fixed points, the ultraviolet and infrared fixed points coalesce in $d=2$, resulting in the clockwise loop flow.
This behavior is reminiscent of the breakdown of the $g$ theorem in the non-Hermitian Kondo effect as well as that of the $c$ theorem in the nonunitary CFT~\cite{fendley1993massless,fendley1993massless2}.  
However, our system is neither quantum impurity (i.e., boundary CFT) nor relativistic CFT, and hence it is not straightforward to find a connection with the breakdown of the $g$ and $c$ theorems.
%\textcolor{red}{[to be modified]}

Following Ref.~\cite{PhysRevLett.121.203001}, we introduce a characteristic scale $t_c$ such that $U_{r,t=t_c}=0$.
Consequently, we find the corresponding energy $E_c=\Lambda_{t_c}^2/2\mu$, that is,
\begin{align}
    E_c%=\frac{\Lambda_0^2}{m}\exp\left(\frac{4\pi g_{r,\Lambda_0}}{m|g_{\Lambda_0}|^2}\right),
    =\frac{\Lambda_0^2}{2\mu}\exp\left(\frac{2U_{r,0}}{|U_{0}|^2}\right),
\end{align}
indicating a non-perturbative dependence on $U_{0}$ as in the case of the non-Hermitian Kondo effect~\cite{PhysRevLett.121.203001}. 
Later, we will show how $E_{\rm c}$ is related to the exact solution.

\section{Two-body resonance and quantum scale anomaly in a non-Hermitian system}
\label{sec:5}

To gain physical insights into the non-Hermitian RG flow in Fig.~\ref{fig:2},
we investigate a non-Hermitian two-body problem that can be solved exactly.
Here, we focus on a two-dimensional system.
The solutions in $d = 1, 3$ are
summarized in Appendix D. 
For $d$=2, an anomalous two-body state appears in a Hermitian system due to the quantum scale anomaly~\cite{PhysRevLett.108.185303}.
In a Hermitian two-body problem, 
from the pole of the two-body $T$ matrix
\begin{align}
\label{eq:30}
    \left[T(E)\right]^{-1}\equiv\frac{S_d\Lambda^{d-2}}{(2\pi)^dU_0}-\Pi(E)=0,
\end{align}
we obtain the two-body binding energy $E_{\rm b}$:
\begin{align}
\label{eq:31}
    E=-E_{\rm b}\equiv-\frac{\Lambda_0^2}{2\mu}\exp\left({\frac{2}{U_{r,0}}}\right).
\end{align}
We see that the expression of $E_{\rm b}$ takes a non-perturbative form for $U_{r,0}$ as in the cases of the Kondo temperature and the BCS pairing gap.
For the non-Hermitian case,
we find the pole in the complex energy plane (i.e., resonance) as $E=E_{\rm R}-i\Gamma$ where $E_{\rm R}$ and $\Gamma$ are the resonance energy and the inverse lifetime, respectively.
Substituting this in Eq.~\eqref{eq:30}, we have
\begin{align}
\label{eq:6}
    1
    &=\frac{U_{0}}{2}\left[\ln\left|\frac{2\mu\sqrt{E_{\rm R}^2+\Gamma^2}}{\Lambda^2}\right|+i{\rm Arg}(-E_{\rm R}+i\Gamma)\right],
\end{align}
leading to
\begin{align}
\label{eq:23}
    E_{\rm R}-i\Gamma%=-\frac{\Lambda_0^2}{m}\exp\left[\frac{4\pi (g_{r,\Lambda_0}+ig_{i,\Lambda_0})}{m|g_{\Lambda_0}|^2}\right].
    =-\frac{\Lambda_0^2}{2\mu}\exp\left(\frac{2}{U_0}\right).
\end{align}
where we use $U_0=\frac{\mu g_{\Lambda_0}}{\pi}$ which is valid for $d=2$.
We find $|E_{\rm R}-i\Gamma|=E_c$, indicating that the characteristic looped RG flow indeed captures the emergence of the resonance state as a consequence of the non-Hermitian quantum scale anomaly.
Note that only the solution with $\Gamma\geq 0$ is allowed because of the relation
${\rm Arg}(-E_{\rm R}+i\Gamma)
=\frac{2U_{i,0}}{|U_{0}|^2}\geq 0
%=\frac{4\pi}{m|g_{\Lambda_0}|^2}g_{i,\Lambda_0}\geq 0
$
obtained from Eq.~\eqref{eq:6}.

\begin{figure}[tb]
    \centering
    \includegraphics[width=0.9\linewidth]{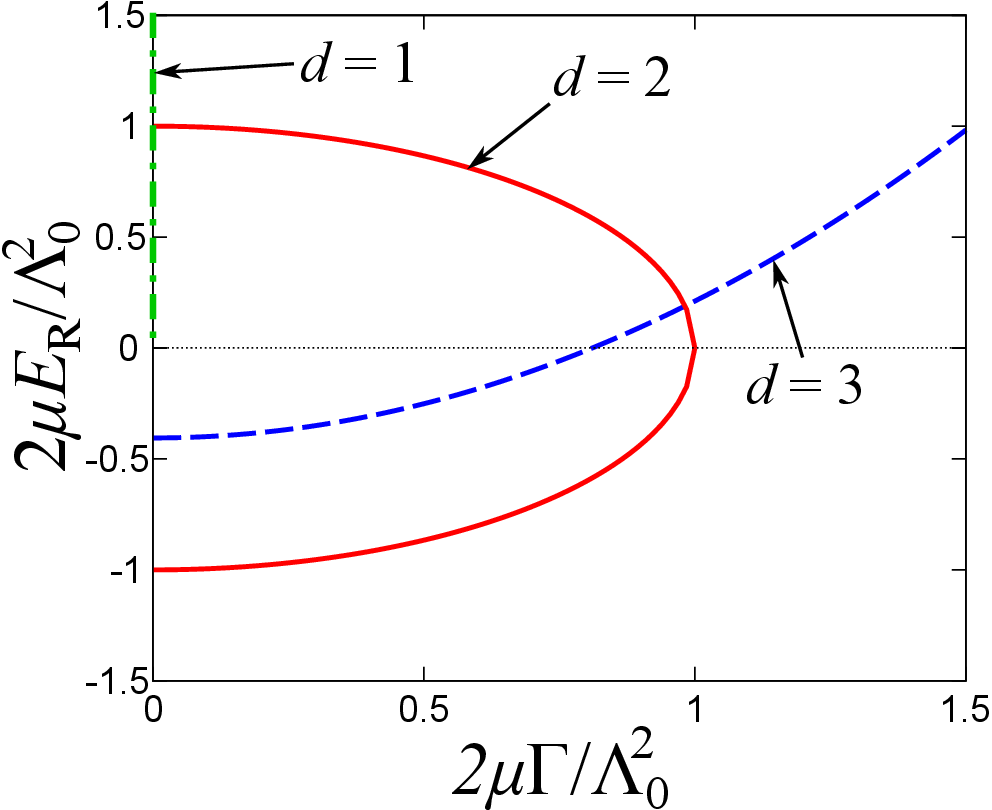}
    \caption{Two-body energy $E=E_{\rm R}-i\Gamma$
    under the pure imaginary two-body coupling $U_0=-iU_{i,0}$
    in the complex plane for spatial dimensions of $d=1,2,3$.}
    \label{fig:3}
\end{figure}

Figure~\ref{fig:3} shows the solutions of $E_{\rm R}$ and $\Gamma$ for a pure imaginary coupling $U_0=-iU_{i,0}$, corresponding to the nonunitary time evolution of a non-interacting two-body system under the postselection of two-particle detection.
Strikingly, in $d=2$, the complex resonance pole forms a semi-circle structure by changing $U_{i,0}$, in contrast to the other dimensions.
In $d=2$, the limit of $U_{i,0}\rightarrow 0$ is not well-defined, reflecting the non-perturbative feature of the non-Hermitian quantum scale anomaly. On the other hand, in the limit of $U_{i,0}\rightarrow \infty$, the energy pole approaches a negative real value $-\Lambda_0^2/2\mu$, that is, a two-body bound state, reminiscent of the quantum Zeno effect~\cite{koshino2005quantum,syassen2008strong}.
In other words, the nonunitary evolution indicates that the information about the unobserved two-particle loss leads to the formation of the bound state.
Such a bound-state formation induced by the Bayesian inference can also be found in $d=3$ ($2\mu E/\Lambda_0^2=-\pi^2/4$ at $U_{i,0}\rightarrow\infty$), whereas the semi-circle behavior of the complex energy pole is absent.
Meanwhile, in $d=1$, the two-body energy always remains a positive real value (i.e., $2\mu E/\Lambda_0^2=\pi^2U_{i,0}^2/4\geq 0$), corresponding to the scattering state.
In this regard, in $d=1$, the Bayesian inference term does not induce any localized two-body state.

The $U_{i,0}$ dependence of the two-body energy in different spatial dimensions can be understood from the non-Hermitian RG flow starting from the real axis in Fig.~\ref{fig:2}.
In $d=2$ and $d=3$, the complex RG flow approaches the origin (i.e., the free limit) at $t\rightarrow\infty$, indicating that a given resonance energy can be induced by an infinitesimally weak imaginary coupling in the infrared limit.
In the limit of $U_{i,0}\rightarrow\infty$,
one can find an infinitely large RG loop in $d=2$ and its deformed one in $d=3$, both of which take an infinitely long RG step to reach the infrared free limit, resulting in the two-body bound state due to the Bayesian inference. 
On the other hand, in $d=1$, the RG flow starting from $U_{r,0}=0$ eventually approaches the repulsive fixed point $U_{r,t\rightarrow\infty}\rightarrow 1$ (corresponding to a Tonks-Girardeau gas~\cite{PhysRevA.75.033608}), indicating the absence of bound and resonance states.

\begin{figure}[tb]
    \centering
    \includegraphics[width=1\linewidth]{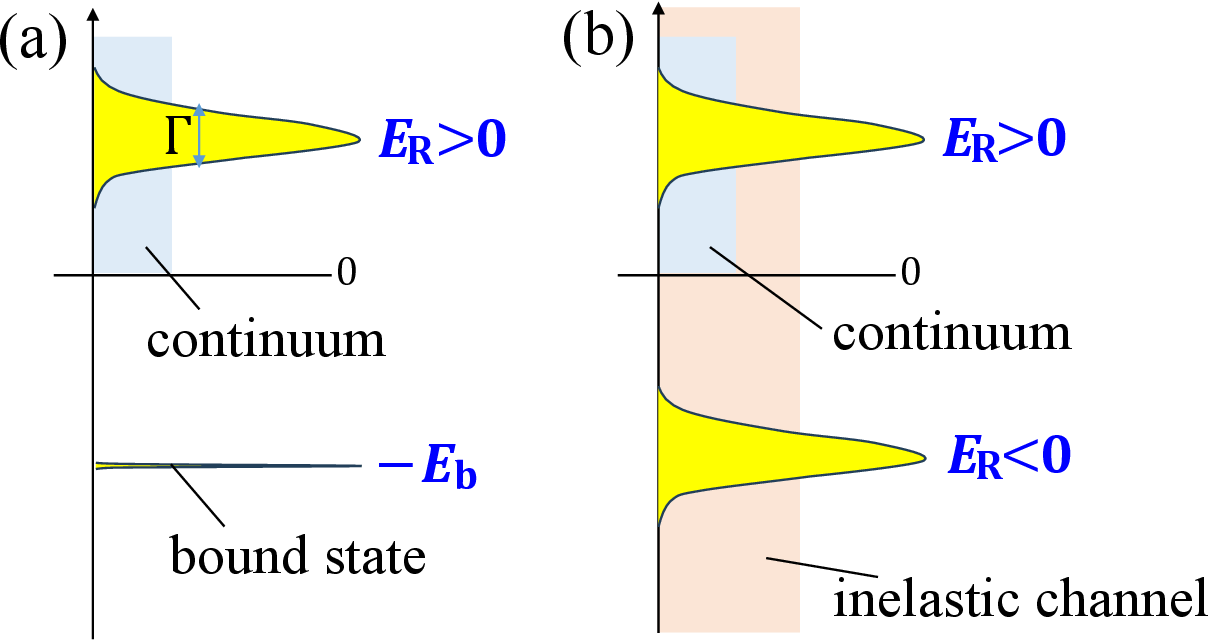}
    \caption{Schematic comparison between (a) the conventional resonance that decays into the continuum and (b) the present case with inelastic loss. In case (b), there are possibilities of both $E_{\rm R}>0$ and $E_{\rm R}<0$.}
    \label{fig:4}
\end{figure}

While the ordinary resonant state has a positive real energy with a nonzero imaginary part by definition, the negative-energy state can also appear as a resonance in the present case.
This difference originates from the distinct decay processes. The conventional resonant state in a Hermitian system decays to the continuum at a positive energy.
On the other hand, even the negative-energy state can decay due to two-body loss in the present non-Hermitian system, as schematically illustrated in Fig.~\ref{fig:4}.
In this regard, the resonance state in the present non-Hermitian system is consistent with the conventional one by shifting the origin of the energy of the environment.

Although the present resonant state decays as $e^{-\Gamma t}$, it is spatially localized in the relative distance $r$ since the two-body wave function in $d=2$ is proportional to the modified Bessel function of the second kind $K_0(r/\xi)$ with a complex-valued localization length 
\begin{align}
\label{eq:34}
    \xi\equiv \xi_r+i\xi_i=
    \frac{1}{\sqrt{-2\mu(E_{\rm R}-i\Gamma)}},
\end{align}
where we introduce $\xi_{r}={\rm Re}[\xi]$ and $\xi_{i}={\rm Im}[\xi]$.
Note that, for $\Gamma>0$, one can find
\begin{align}
\xi_r=\frac{1}{\sqrt{2\mu\sqrt{E_{\rm R}^2+\Gamma^2}}}\cos\left(\frac{\phi+\pi}{2}\right)>0,    
\end{align}
where $\phi={\rm Arg}(E_{\rm R}-i\Gamma)$ and $-\pi\leq\phi<0$.
In this regard, the two-body wave function of the resonant state exponentially decreases at a long distance as $\sim e^{-r/\xi_{r}}/\sqrt{r}$ (see also Appendix~\ref{app:e}).
For the bound state with $\phi=-\pi$ (i.e., $E_{\rm R}\rightarrow-E_{\rm b}$ and $\Gamma\rightarrow 0$), $\xi$ is equal to the scattering length $a=\sqrt{2\mu E_{\rm b}}$, resulting in the localized wave function $\sim e^{-r/a}/\sqrt{r}$.
On the other hand, if the energy pole approaches the virtual state with $\phi=0$ (i.e., $E_{\rm R}>0$ and $\Gamma\rightarrow 0$), $\xi_r$ becomes negative and thus the wave function diverges at long distance.
This is similar to the case of a dissipative Kondo effect where the localization length of the Kondo singlet state diverges when the associated Kondo energy scale becomes positive~\cite{PhysRevB.111.125157}.

We note that the present two-body resonant state does not exhibit a divergence at a spatially long distance in contrast to the Gamov state~\cite{Gamow1928}. This difference originates from the fact that the present two-body resonant state decays at a short distance due to the inelastic two-body loss, while the Gamov state describes the decay process toward a largely separated regime. 

\begin{figure*}[t]
    \centering
    \includegraphics[width=1\linewidth]{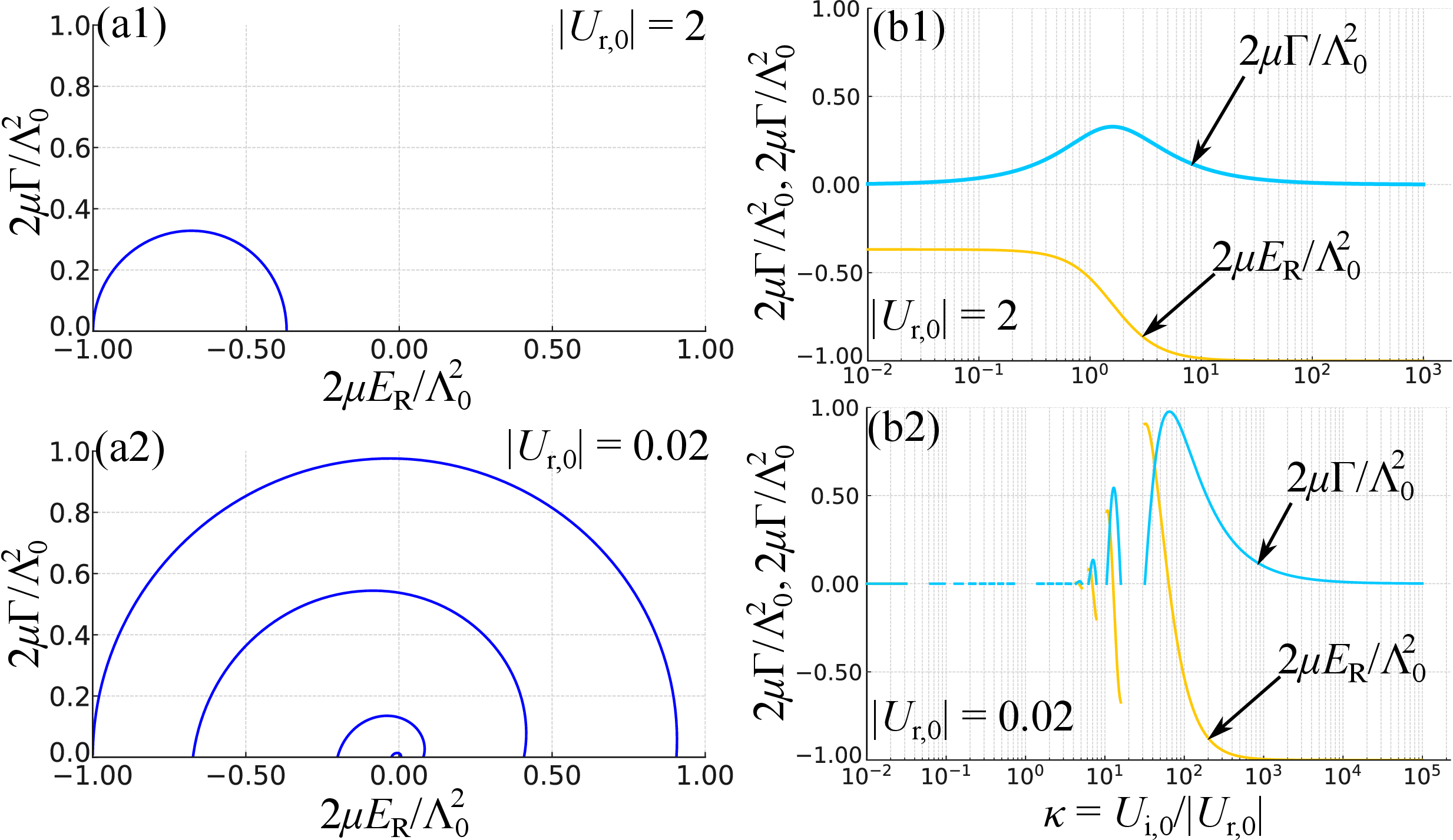}
    \caption{Two-dimensional two-body energy pole $E=E_{\rm R}-i\Gamma$ obtained by changing $\kappa=U_{i,0}/|U_{r,0}|$ at (a1) strong coupling ($|U_{r,0}|=2$) and at (a2) weak coupling ($|U_{r,0}|=0.02$).
    In the panels (b1) and (b2), the imaginary coupling dependence of $E_{\rm R}$ and $\Gamma$ on $\kappa$ are plotted at (b1) $|U_{r,0}|=2$ and (b2) $|U_{r,0}|=0.02$.
    }
    \label{fig:5}
\end{figure*}

Finally, we discuss a non-trivial feature of the non-perturbative resonant state in $d=2$ due to the non-Hermitian quantum scale anomaly.
Figures~\ref{fig:5} (a1) and (a2) show the trajectory of the two-body energy pole in the complex plane of $E_{\rm R}$ and $\Gamma$ at $|U_{r,0}|=2$ and $|U_{r,0}|=0.02$. 
Figures~\ref{fig:5}(b1) and (b2) present their explicit coupling dependencies where $\kappa=U_{i,0}/|U_{r,0}|$ is the dimensionless imaginary coupling.
In the strong-coupling case (i.e., $|U_{r,0}|=2$),
the bound state with $2\mu E/\Lambda_0^2\equiv -e^{-2/|U_0|}=-e^{-1}$ at $\kappa=0$ changes to the resonant state with $\kappa\neq 0$, and eventually the quantum Zeno-like bound state with $2\mu E/\Lambda_0^2=-1$ at $\kappa\rightarrow\infty$.
On the other hand, at weak coupling (i.e., $|U_{r,0}|=0.02$), the rise and fall of two-body resonant states can be found.
At $\kappa=0$, the two-body bound state with $2\mu E/\Lambda_0^2\equiv-e^{-2/|U_0|}=-e^{-100}$ is extremely shallow, and virtually disappears at small $\kappa$.
However, at large $\kappa$, another resonant state appears again at a positive energy
This second resonant state becomes a bound state and disappears again at larger $\kappa$; such appearance and disappearance of a bound state repeat with increasing $\kappa$.
The number $N_{\rm ER}$ of emergent resonant states is given by
\begin{align}
    N_{\rm ER}(U_{r,0})=2\left\lfloor\frac{\lfloor(\pi|U_{r,0}|)^{-1}\rfloor+1}{2}\right\rfloor-\delta N,
\end{align}
where $\lfloor x\rfloor$ is the floor function, and $\delta N=1$ if $1/\pi |U_{r,0}|$ is an odd integer (and $\delta N=0$ otherwise).
In addition to the emergent resonant state, one may find the conventional bound state at $\kappa=0$ and the quantum-Zeno-like bound state at $\kappa\rightarrow\infty$.
Indeed, while we only find the bound states at $\kappa=0$ and at $\kappa\rightarrow \infty$ without emergent resonant states in the case of $|U_{r,0}|=2$, there are $16$ emergent resonant states in the case of $|U_{r,0}|=0.02$. (Those with smaller $|E_{\rm R}|$ are not shown in Fig.~\ref{fig:5} because they are too small in the present energy scale.)

For experimental observations, the non-Hermitian quantum scale anomaly can be detected as the appearance and disappearance of localized two-body states by tuning the two-body loss rate under a two-dimensional confinement. In an ultracold atom experiment, the two-body loss and the spatial dimensions can be controlled by the photoassociation process~\cite{RevModPhys.78.483} and the external trap potential~\cite{PhysRevLett.105.030404}, respectively. The localized wave function can be observed by the matter-wave spectroscopy with the recent magnification technique~\cite{wdjr-m2hg}.
It is also feasible to observe the appearance and disappearance of the complex eigenenergy via absorption spectroscopy~\cite{PhysRevLett.129.093001}.

While the non-Hermitian quantum scale anomaly in $d=2$ is relevant to cold atomic systems, the two-dimensional system might be approximately realized in alpha particles near the surface of neutron-rich nuclei~\cite{tanaka2021formation}.
Since alpha particles may feel an imaginary potential associated with the formation of light nuclei as a subsystem, such as $^{12}$C and $^{16}$O~\cite{PhysRevLett.87.192501}, the measurement backaction leads to a localized configuration of alpha particles.
These can be probed by measuring the cluster decay process~\cite{PhysRevC.51.594,santhosh2002exotic}.
Such a possibility opens a fascinating direction in the research of radioactive heavy nuclei.

\section{Coherent neutron-nucleus scattering in $d=3$}
\label{sec:6}
In nuclear physics, we consider a coherent neutron-nucleus scattering in $d=3$~\cite{sears1992neutron}.
The $T$ matrix pole is expressed in terms of $a$ as~\cite{PhysRevResearch.5.043010}
\begin{align}
    {E}_{\rm R}-i\Gamma%&=-\frac{(a_r-ia_i)^2}{2\mu |a|^2}\cr
    &=\frac{a_i^2-a_r^2}{2\mu |a|^4}+i\frac{a_r a_i}{\mu|a|^4}
\end{align}
where $a=a_r+ia_i$, and $a_r a_i<0$ is required for a physically relevant resonance.
Indeed, one can find 
\begin{align}
    a_i
 &=-\frac{\pi |U_0|^2}{2\Lambda_0}\frac{U_{i,0}}{(U_{r,0}+|U_0|^2)^2+U_{i,0}^2}<0.
\end{align}
Since nonzero $a_i$ is induced by $U_{i,0}$ under the conditional selection for the elastic channel (i.e., without inelastic loss), it can be interpreted as an example of the measurement backaction.
In turn, 
\begin{align}
    a_r
           &=\frac{\pi |U_0|^2}{2\Lambda_0}\frac{U_{r,0}+|U_0|^2}{(U_{r,0}+|U_0|^2)^2+U_{i,0}^2}
\end{align}
can also be negative when $U_{r,0}$ and $U_{i,0}$ are sufficiently small such that
\begin{align}
\label{eq:36}
    %U_{r,0}^2+U_{r,0}+U_{i,0}^2<0
    \left(U_{r,0}+\frac{1}{2}\right)^2+U_{i,0}^2<\frac{1}{4}.
\end{align}
In this regard, the state with $a_ra_i>0$ correspond to the RG flow near the line of $U_{i,t}=0$ from $U_0=-1$ to $U_0=0$ in Fig.~\ref{fig:2}(c).
In analogy with the unitary point $U_{0}=-1$ in the Hermitian case,
Eq.~\eqref{eq:36} can be regarded as a \textit{critical semicircle}
where the sign of $a_r$ is changed and hence there is a threshold for the resonant state.
On the critical semicircle, $a$ does not diverge in the presence of nonzero $U_{i,0}$.
Instead, we find $a_r=0$ and $a_i=\frac{\pi |U_0|^2}{2\Lambda_0 U_{i,0}}$.
Remarkably, 
in Fig.~\ref{fig:2}(c), there is an RG flow 
running from the unitary fixed point ($U_{0}=-1$) to the free fixed point ($U_{t\rightarrow\infty}=0$)
along the critical semicircle.
In this sense, it can be understood that the critical semicircle is a boundary between
 the resonant flow ($\frac{\partial U_{r,t}}{\partial t}<0$ in the ultraviolet regime around $t=0$) and the non-resonant flow with $\frac{\partial U_{r,t}}{\partial t}>0$ for any $t$.

Let us estimate the complex scattering length from the cross section~\cite{sears1992neutron}.
Using Eq.~\eqref{eq:sigma_tot}, we obtain the total cross section
\begin{align}
    \sigma_{\rm tot}%&=
   % \frac{4\pi}{k}{\rm Im}\left[\frac{1}{-1/a-ik}\right]\cr
   % &=-\frac{4\pi|a|^2}{k}{\rm Im}\left[\frac{1}{a_r-i(a_i-k|a|^2)}\right]\cr
    &=4\pi|a|^2\frac{|a|^2-a_i/k}{a_r^2+(k|a|^2-a_i)^2}.
\end{align}
In this regard, $\sigma_{\rm tot}$ exhibits infrared divergence as $-a_i/k$ due to the absorption represented by the imaginary part of the scattering length.
The absorption cross section $\sigma_{\rm abs}$ is given by
\begin{align}
    \sigma_{\rm abs}&=\sigma_{\rm tot}-\sigma_{\rm el}\cr
%    &=4\pi|a|^2\frac{|a|^2-a_i/k}{a_r^2+(k|a|^2-a_i)^2}
%    -4\pi\frac{|a|^4}{|a_r^2+(a_i-k|a|^2)^2}\cr
    &=-\frac{4\pi a_i}{k}\frac{|a|^2}{a_r^2+(k|a|^2-a_i)^2},
\end{align}
where 
\begin{align}
\sigma_{\rm el}=\frac{4\pi}{|1/a+ik|^2}=   \frac{4\pi|a|^4}{a_r^2+(a_i-k|a|^2)^2} 
\end{align}
is the elastic cross section.
At small $k$, we obtain $a_i\simeq -k\sigma_{\rm abs}/4\pi$.
In Table~\ref{tab:bc},
we list the estimated $a_r$ and $a_i$ for several nuclei.
The zero-range approximation may be valid when
$|a|$ is large compared to the typical nuclear radius $\sim 6\,{\rm fm}$. 
\begin{table}[bt]
\begin{center}
\caption{Coherent neutron-nucleus scattering lengths of several nuclei obtained from Ref.~\cite{sears1992neutron}.}
\label{tab:bc}
  \begin{ruledtabular}
\begin{tabular}{c|c|c}
%\hline
     & $a_r$ [fm] & $a_i$ [fm] \\
    \hline 
    $^{155}$Gd & $6.0$ & $-17$  \\
        $^{157}$Gd & $-1.14$ & $-72$  \\
                $^{168}$Yb & $-4.07$ & $-0.62$  \\
                        $^{196}$Hg & $30.3$ & $-0.86$  \\
 %   \\ \hline
\end{tabular}
\end{ruledtabular}
\end{center}
\end{table}

Moreover, one can obtain the dimensionless complex-valued coupling for a given cutoff $\Lambda_t$ as
\begin{align}
\label{eq:40}
    U_t=\frac{1}{\frac{\pi}{2\Lambda_t a}-1}.
\end{align}
Equation~\eqref{eq:40} indicates that the RG flow orbit in Fig.~\ref{fig:2}(c) is governed by $a$.
It is easily found that Eq.~\eqref{eq:40} reproduces the fixed points $U_{t=0}=-1$ and $U_{t\rightarrow \infty}=0$ at $\Lambda_0\rightarrow \infty$ and $\Lambda_\infty\rightarrow0$, respectively.

\begin{figure}[t]
    \centering
    \includegraphics[width=1\linewidth]{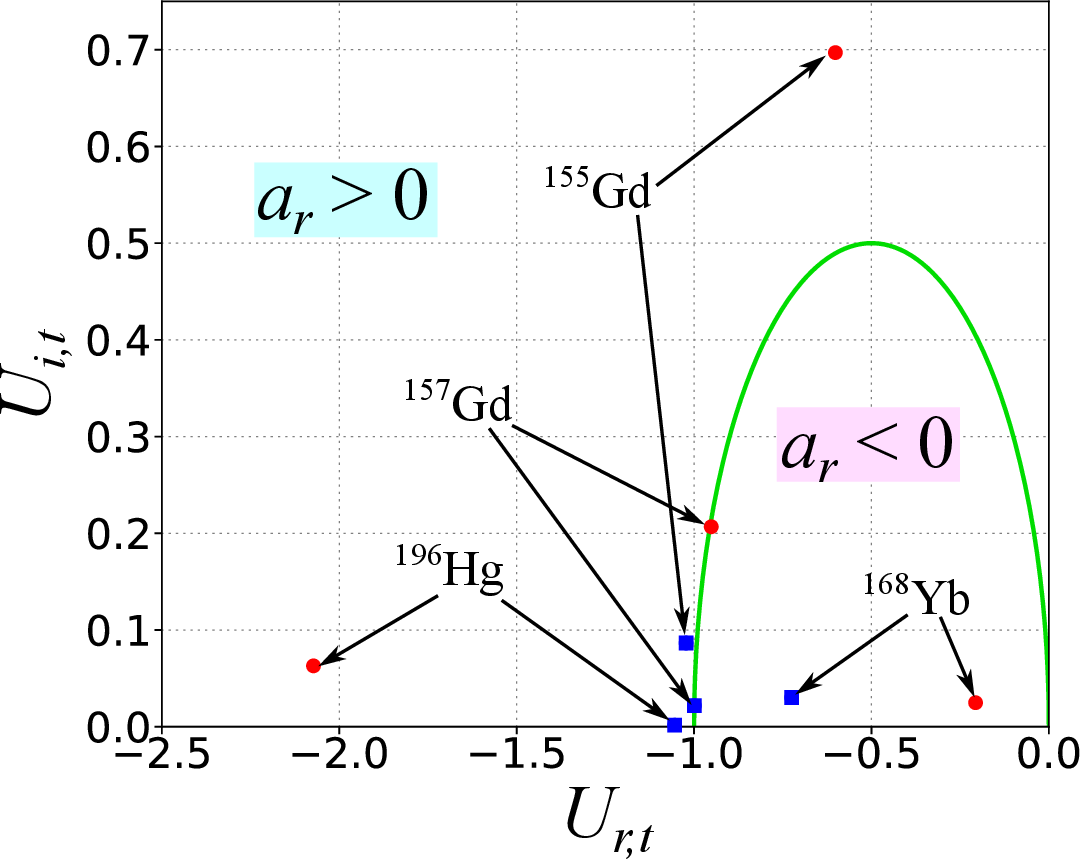}
    \caption{Phase diagram of a neutron-nucleus system in the plane of $U_{r,t}$ and $U_{i,t}$. The green solid curve indicates the critical semicircle given by Eq.~\eqref{eq:36}, where the sign of $a_r$ is changed. We also plot the data of several nuclei ($^{155}$Gd, $^{157}$Gd, $^{168}$Yb, and $^{196}$Hg)~\cite{sears1992neutron}, where the red cicles and the blue square show the cases of $\Lambda_t=0.1\,{\rm fm}^{-1}$ and $\Lambda_t=1.0\,{\rm fm}^{-1}$, respectively.
    Note that $U_{t}=-1$ represents the unitary point in the Hermitian case.
    }
    \label{fig:6}
\end{figure}

Figure~\ref{fig:6} represents the region of $a_{r}<0$ enclosed by the critical semicircle
in the plane of the dimensionless coupling $U_{t}=U_{r,t}-iU_{i,t}$.
For comparison, we plot the coupling constants estimated from the cross section data~\cite{sears1992neutron}.
Since we consider a slow incident neutron to ensure that the finite-range effect is negligible,
we take $\Lambda_t\simeq 0.1\sim 1.0 \,{\rm fm}^{-1}$ which could be of the order of the inverse effective range.
While these points approach the unitary point $U_{t}=-1$ with increasing $\Lambda_t$ in Eq.~\eqref{eq:40},
they never cross the critical semicircle (e.g., $^{168}$Yb with $a_r=-4.07\,{\rm fm}$ is always inside of the critical semicircle).
In particular, the data of $^{157}$Gd, which is known as a strong neutron absorber, are close to the critical semicircle even for different $\Lambda$.
This result suggests a new type of universal states in non-Hermitian two-body systems.
However, it should be noted that the scattering length and the cross section in strong neutron absorbers are strongly energy-dependent.
A further investigation will be left for future work.

\section{Dineutron resonance in Borromean two-neutron halo nuclei}
\label{sec:7}
In this section, we discuss the relationship between the quantum measurement effect and the localized dineutron correlations in a two-neutron halo nucleus.
First, we consider an isolated two-neutron system.
Within the contact-type two-body potential, the coupling constant $g_{nn}\in\mathbb{R}$ is given by
\begin{align}
    \frac{\mu}{2\pi a_{nn}}-\frac{\mu\Lambda_{nn}}{\pi^2}=\frac{1}{g_{nn} }<0,
\end{align}
where $a_{nn}=-18.5~{\rm fm}$ is the $^1S_0$ neutron-neutron scattering length~\cite{PhysRevC.51.38}.
For the two-neutron systems, we get $\mu=M_n/2$ where $M_n=939~{\rm MeV}$  is the neutron mass.
In the present case, the momentum cutoff $\Lambda_{nn}$ can be associated with the effective range $r_{nn}=2.7~{\rm fm}$ of the neutron-neutron interaction as $\Lambda_{nn}=4/\pi r_{nn}$~\cite{ohashi2020bcs}.
In this case,
the system does not have a bound state and only a virtual state with $E=-1/M_na_{nn}^2$, where the two-body wave function diverges at a large distance as $\sim e^{r/|a_{nn}|}/r$.

\begin{figure}[t]
    \centering
    \includegraphics[width=1\linewidth]{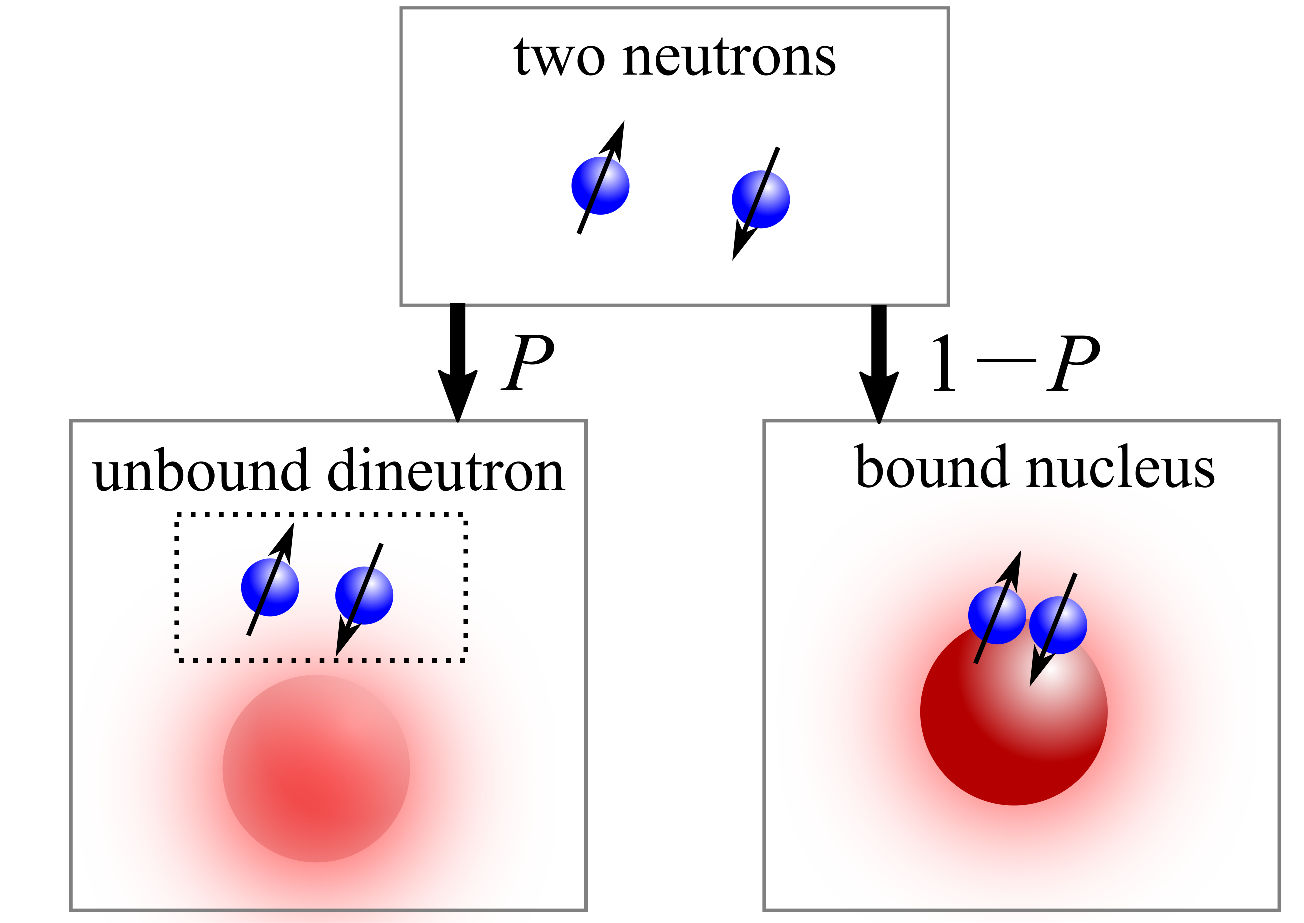}
    \caption{Schematic illustration of the dineutron subsystem in two-neutron halo nuclei. The dineutron subsystem can be projected by the projector $P$, while the bound nucleus can be regarded as the null state of neutrons, and taken out by the projector $1-P$. }
    \label{fig:7}
\end{figure}

Then, we consider an additional imaginary coupling induced by the absorption of two neutrons to the core nucleus because of the formation of the bound Borromean nuclei, as shown in Fig.~\ref{fig:7}.
By identifying the absorption process as a two-body loss, the effective interaction strength is given by
\begin{align}
    G_{nn}=g_{nn}-iu,
\end{align}
where $u>0$ is a real value.
In principle, $u$ can be obtained by the Feshbach projection formalism~\cite{feshbach1958unified,feshbach1962unified}. 
In this paper, we treat $u$ as a free parameter to understand the role of $u$ qualitatively.
Since we are interested in Borromean nuclei where any two-body subsystems are unbound, the one-body loss related to the formation of two-body bound states can be neglected. 
The pole of the $T$ matrix leads to the dineutron energy
\begin{align}
    E=-\frac{1}{M_na_{\rm eff}^2},
\end{align}
where we introduce an effective scattering length $a_{\rm eff}$ 
which is determined so as to satisfy
\begin{align}
\label{eq:48}
     \frac{|a_{nn}|}{ a_{\rm eff}}&=-1+
    \left(1+\frac{8 |a_{nn}|}{\pi^2 r_{nn}}\right)
    \frac{x^2}{1+x^2}\cr
    &\quad +i\left(1+\frac{8 |a_{nn}|}{\pi^2 r_{nn}}\right)
    \frac{x}{1+x^2},
\end{align}
with a dimensionless imaginary coupling strength $x=u/|g_{nn}|$.

\begin{figure}[t]
    \centering
    \includegraphics[width=0.9\linewidth]{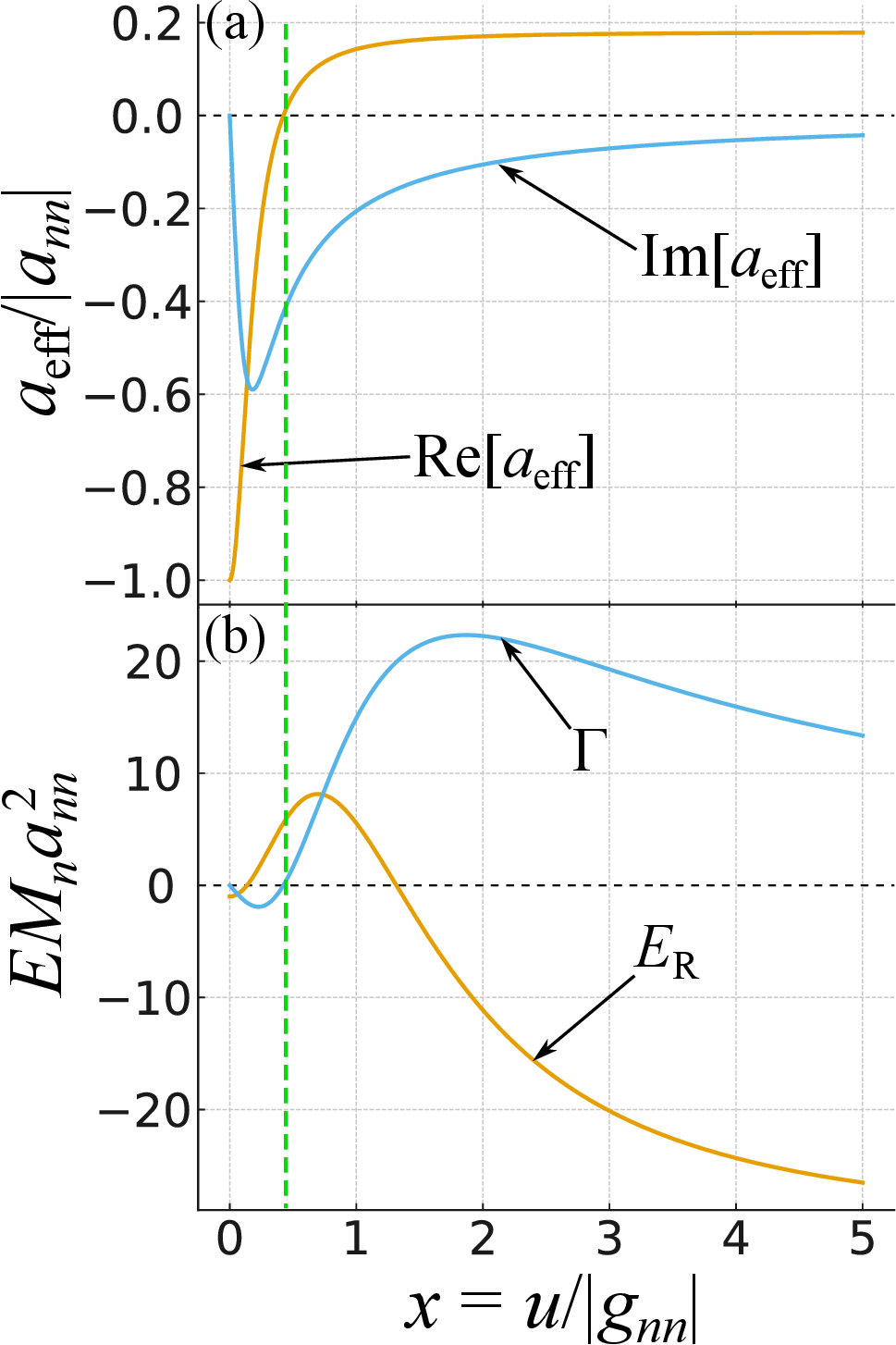}
    \caption{(a) Effective scattering length $a_{\rm eff}/|a_{nn}|$ and (b) two-body energy $EM_na_{nn}^2$
    in a dineutron subsystem of Borromean two-neutron halo nuclei 
    as functions of the imaginary coupling constant $x=u/|g_{nn}|$ representing the absorption of two neutrons into the core nucleus (see also Eq.~\eqref{eq:48}). The dashed vertical line represents the critical semicircle $x\simeq 0.42$ characterized by the sign change of $a_{\rm eff}$.}
    \label{fig:8}
\end{figure}

Figure~\ref{fig:8} shows $a_{\rm eff}/|a_{nn}|$ and $EM_na_{nn}^2=(E_{\rm R}-i\Gamma)M_na_{nn}^2$ as functions of $x$.
At $x=0$, we find $a_{\rm eff}=-|a_{nn}|\equiv a_{nn}$ and $E=-1/M_na_{nn}^2$, corresponding to the virtual state in free space.
For small but nonzero $x$, $a_{\rm eff}<0$ and $\Gamma<0$ indicate that the localized resonant dineutrons are absent up to $x\simeq 0.42$.
Based on the discussion in Sec.~\ref{sec:6},
$a_{\rm eff}=0$ corresponds to the critical semicircle.
Beyond this, $a_{\rm eff}$ becomes positive, and a resonant state appears with its localization length given by $\xi_r={\rm Re}[a_{\rm eff}]$, where the two-body wave function takes the form of $e^{-r/\xi_r}/r$.
Eventually, at $x\rightarrow\infty$, we find a localized dineutron with $EM_na_{nn}^2=-64a_{nn}^2/\pi^4r_{nn}^2\simeq -30.8$ and $\xi_r=\pi^2r_{nn}/8\simeq 3.3~{\rm fm}$.
The peaked behavior of $\Gamma$ in Fig.~\ref{fig:8}(a) is due to an interplay between $g_{nn}$ and $u$,
where $g_{nn}$ plays a crucial role for crossing the critical semicircle from the virtual state to the resonant state at $x\lesssim 1$ and the effects of $u$ becomes dominant at $x\gesim 1$ as $E_{\rm R}\rightarrow-{4\Lambda_{nn}^2}/{\pi^2 M_n}$ and $\Gamma\rightarrow 0$ in the large pure imaginary coupling limit (see also Fig.~\ref{fig:3}).

Remarkably, the upper bound $\xi_r\simeq 3.3~{\rm fm}$ is close to the relative two-neutron distance in the three-body analysis~\cite{PhysRevC.72.044321}, regardless of our simplified calculation and the fact that we do not specify the core nucleus.
This universal feature, in the sense that the localization length does not depend explicitly on the neutron-core interaction, shares a similarity with the recent field-theoretical description of Borromean halo nuclei~\cite{PhysRevLett.128.212501}.
On the other hand, while our calculation corresponds to dineutrons in the excited nuclei,
the three-body analysis was performed for the ground state~\cite{PhysRevC.72.044321}.
In this sense, our prediction might be more relevant to the strong dineutron correlation of the excited
$0_2^+$ state of $^8$He~\cite{PhysRevLett.131.242501}, where the compact dineutron size $\sim2\,{\rm fm}$ has been reported in comparison with a cluster model calculation.
Provided that $-E_{\rm R}$ in Fig.~\ref{fig:8}(b)
is similar to the two-neutron separation energy $S_{2n}$ (where $S_{2n}\simeq 1\sim 2~{\rm MeV}$ for $^{6,8}$He~\cite{RevModPhys.76.215}),
the corresponding imaginary potential reads $x\simeq 1.4$ from Fig.~\ref{fig:8}(b), resulting in the localized dineutron state with $\xi_{r}\simeq 3.0~{\rm fm}$ that is qualitatively consistent with the cluster model calculation~\cite{PhysRevLett.131.242501}.

Our result gives a new insight into multi-neutron correlations in neutron-rich nuclei.
From our analysis, it is clear how a two-neutron system is modified by the presence of the two-neutron absorber, that is, the core nucleus.
It would be reasonable to consider similar effects on the observation of a tetraneutron in
the double-charge reaction~\cite{PhysRevLett.116.052501} and
the $\alpha$-knockout reaction of $^8$He~\cite{duer2022observation}.
While this experiment reports the resonant tetraneutron state,
a few-body theoretical analysis indicates the absence of a compact four-neutron state~\cite{marques2021quest}.
Once a multi-neutron absorber (i.e., $\alpha$ particle in the case of Ref.~\cite{marques2021quest}) exists,
there is a possibility of observing a localized resonant state that is absent in an isolated multi-neutron system, as a result of the quantum-measurement effect where dineutron subsystems without undergoing the jump process (absorption to the core nuclei) show the required nonunitary state change.
The non-Hermitian description associated with the absorption process can also be useful for a multi-neutron subsystem in unstable neutron-rich nuclei such as $^{26}$O~\cite{PhysRevLett.116.102503} and $^{28}$O~\cite{kondo2023first}.

\section{Summary}
\label{sec:8}
In this paper, we have developed the non-Hermitian RG approach in strongly interacting few-body systems with inelastic collisions, to build a bridge between AMO physics and nuclear physics from a few-body perspective.
We have shown that the RG equation can be obtained from the condition that the two-body scattering $T$ matrix is invariant under the RG transformation in the presence of two-body loss. 
As the information about the absence of the quantum jump process leads to the nonunitary time evolution of a continuously monitored system,
the RG transformation involves the complex-valued running coupling constant, in contrast to an isolated quantum system.
These changes in the time evolution and in the RG flow originate from the Bayesian inference as a consequence of the update of the observer's knowledge.
From a viewpoint of nuclear physics, 
the resonant state observed in high-energy experiments can also be interpreted as due to the backaction of quantum measurement.

We have demonstrated our approach to a non-relativistic two-body system that can be analytically solved and can be verified in AMO and nuclear experiments.
We thus obtain a clear physical correspondence between the non-Hermitian RG flow and the exact solution.
In particular, in two dimensions where non-perturbative effects such as quantum scale anomaly and asymptotic freedom appear,
we have shown that the RG flow exhibits a loop structure in the complex plane of the running coupling constant due to the collision of ultraviolet and infrared fixed points.
The characteristic energy scale for the looped RG flow structure is found to be consistent with the two-body resonance energy.
%We have also shown that the non-Hermitian exact RG equation is equivalent to that obtained by the Wilsonian RG approach in a two-body problem.

To gain physical insights into non-trivial RG flow and the consequences of non-Hermitian quantum scale anomaly, we have explored the fate of a dissipative two-body state and shown that a looped RG flow represents the non-Hermitian quantum scale anomaly which induces the appearance and disappearance of non-perturbative resonant states with increasing the non-Hermitian interaction corresponding to the strength of inelastic collision.
For pure imaginary coupling, the measurement-induced resonant state can be found in $d=2$ and $d=3$, where the non-Hermitian RG flow approaches the free fixed point in the infrared limit, in contrast to $d=1$ with the repulsive infrared fixed point.

To apply our framework to nuclear systems,
we have examined the coherent neutron-nucleus scattering where the scattering length and the cross section are summarized in Ref.~\cite{sears1992neutron}.
On the basis of the complex scattering length in $d=3$ which results from the conditional measurement backaction,
we have generalize the concept of the unitary point in the Hermitian system to the critical semicircle in the non-Hermitian system, which corresponds to the threshold of the resonant state obtained from the $T$ matrix pole.
It is found that $^{155}$Gd and $^{157}$Gd, which are known as strong neutron absorbers, are close to the critical semicircle regardless of their large imaginary part of the scattering length.
We have demonstrated that the non-Hermitian RG flow orbit is characterized by the complex scattering length.

Finally, we have shown that the strong dineutron correlations in two-neutron Borromean halo nuclei can be understood as a consequence of the quantum measurement effect, where the dineutron subsystem undergoes a nonunitary state change even without the quantum jump process due to the absorption into the core nucleus. 
Our result of the dineutron localization length is close to that obtained in the three-body analysis~\cite{PhysRevC.72.044321} and the cluster model calculation~\cite{PhysRevLett.131.242501}, indicating that the dineutron is shrunk by the measurement backaction.
Our scenario can be tested by comparison with the core nuclei 
which may exhibit different dineutron absorption strengths.

For future perspectives, it is worth investigating other non-Hermitian few-body systems, such as a two-body system with the $p$-wave interaction and the finite-range interaction, and three- and four-body systems.
Moreover, the few-body open-system description of the subsystems in complicated nuclear structures (e.g., multineutron states in halo nuclei) may open a new perspective on quantum measurement effects in nuclear experiments.

\acknowledgments
The authors thank Hongchao Li, Tingyu Zhang, Daisuke Suzuki, and Sibo Wang for the useful discussion.
H.~T. was supported by JSPS KAKENHI for Grants (Nos.~JP22K13981, JP23K22429) from MEXT, Japan.
M.~N. was supported by JSPS KAKENHI Grant No. JP24K16989. M.~U. was supported by JSPS KAKENHI Grant No. JP22H01152 and the CREST program “Quantum Frontiers” of JST (Grand No. JPMJCR23I1).
\appendix

\section{Non-Hermitian Schr\"{o}dinger equation and scattering $T$ matrix}
\label{app:a}
To obtain a quantum state $|\Psi(t)\rangle$ in the scattering problem under the nonunitary evolution governed by a non-Hermitian Hamiltonian $H_{\rm eff}=H_0+V$, we employ the interaction picture
\begin{align}
    |\Psi_{\rm I}(t)\rangle=e^{iH_0t}|\Psi(t)\rangle.
\end{align}
The non-Hermitian Schr\"{o}dinger equation yields
\begin{align}
\label{eq:B3}
    i\frac{d}{d t}|\Psi_{\rm I}(t)\rangle
    &=V_{\rm I}(t)|\Psi_{\rm I}(t)\rangle,
\end{align}
where
    $V_{\rm I}(t)=e^{iH_0t}Ve^{-iH_0t}$ is the interaction representation of $V$ with respect to $H_0$.
At $t=\rightarrow -\infty$, we assume $|\Psi(t\rightarrow-\infty)\rangle=|\Phi_0\rangle$, where $|\Phi_0\rangle$ is the eigenstate of $H_0$ as $H_0|\Phi_0\rangle=E|\Phi_0\rangle$, and $V$ is adiabatically switched on as $V\rightarrow e^{\delta t}V$, where $\delta$ is an infinitesimal positive number.
Formally, we have the solution of Eq.~\eqref{eq:B3} as
\begin{align}
\label{eq:B4}
    |\Psi_{\rm I}(t)\rangle=|\Phi_0\rangle-i\int_{-\infty}^{t}dt' \, V_{\rm I}(t')|\Psi_{\rm I}(t')\rangle.
\end{align}

We are interested in the solution in the form of
\begin{align}
\label{eq:B5}
    |\Psi(t)\rangle=e^{-iEt}|\Psi_0\rangle,
\end{align}
where $|\Psi_0\rangle$ is a stationary wave function.
Note that $E$ is a real-valued on-shell incident energy (i.e., not an eigenenergy of $H$).
Substituting Eq.~\eqref{eq:B5} into Eq.~\eqref{eq:B4},
we obtain the Lippmann-Schwinger equation in Eq.~\eqref{eq:11}.
The $T$-matrix operator is introduced as
\begin{align}
    T(E)|\Phi_0\rangle=V|\Psi_0\rangle,
\end{align}
indicating that the $T$ matrix describes how the outgoing wave function is modified by $V$.
We thus obtain
\begin{align}
    V|\Psi_0\rangle =V|\Phi_0\rangle+V\Pi(E)V|\Psi_0\rangle,
\end{align}
which leads to
\begin{align}
\label{eq:B7}
    T(E)=V+V\Pi(E)T(E),
\end{align}
where $\Pi(E)=1/(E+i\delta-H_0)$ is the two-body propagator.

\section{Renormalization group based on the momentum subtraction}
\label{app:b}
In this section,  we show that the center-of-mass momentum $\bm{P}$ can be used as the probing RG scale in the two-body system.
We consider $d=3$.
In the rest frame, we obtain the $T$ matrix with the running coupling $g_\Lambda\rightarrow g_P$ ($P=|\bm{P}|$) as
\begin{align}
    &T(E-P^2/2M)\cr
    &\quad=\left[\frac{1}{g_P}+\frac{\mu\Lambda_0}{\pi^2}
    +i\frac{\mu\sqrt{2\mu(E-P^2/2M)}}{2\pi}
    \right]^{-1},
\end{align}
where $M$ is the total mass and $-P^2/2M$ is the center-of-mass energy shift.
We impose the invariance of the low-energy $T$ matrix under a change in $P$ as
\begin{align}
    \frac{\partial }{\partial P}T(-P^2/2M)=0,
\end{align}
which leads to
\begin{align}
\label{eq:38}
    \frac{dg_P}{dP}=-\frac{\mu}{2\pi}\sqrt{\frac{\mu}{M}}g_P^2.
\end{align}
Equation \eqref{eq:38} is equivalent to Eq.~\eqref{eq:19} with $d=3$ except for the coefficients which reflect the difference of the RG subtraction schemes.
Indeed, at $E=0$, one can rewrite $T(-P^2/2M)$ as
\begin{align}
    T(-P^2/2M)=\left(\frac{1}{g_P}+\frac{\mu}{\pi^2}\tilde{\Lambda}_P\right)^{-1},
\end{align}
where 
\begin{align}
\tilde{\Lambda}_P=\Lambda_0-\frac{\pi}{2}\sqrt{\frac{\mu}{M}}P.    
\end{align}
Consequently, Eq.~\eqref{eq:38} can also be interpreted as the RG evolution of $\tilde{\Lambda}_P$ from $P=0$ ($\tilde{\Lambda}_{P}=\Lambda_0$) to $P=\frac{2}{\pi}\sqrt{\frac{M}{\mu}}\Lambda_0$ ($\tilde{\Lambda}_P=0$).
The corresponding RG equation reads $\frac{dg_P}{d\tilde{\Lambda}_P}=\frac{\mu}{\pi^2}g_P^2$.

\section{Wilsonian perturbative renormalization group in a non-Hermitian two-body system}
\label{app:c}
To make this paper self-contained, we derive the non-Hermitian RG equation
in a non-relativistic two-body system on the basis of the Wilsonian perturbative RG method~\cite{RevModPhys.66.129}.
An effective Hamiltonian can be separated into the subspaces of slow and fast modes as
\begin{align}
    H_{\rm eff}=H_{>}+H_{<}+V_{\rm ex},
\end{align}
where $V_{\rm ex}$ represents the interaction between fast and slow modes.
The partition function is given by
%\begin{align}
    $Z={\rm Tr}\left[e^{-\beta H}\right]$,
%\end{align}
where ${\rm Tr}\,[\cdots]$ denotes the trace in the entire Hilbert space.
Here the inverse temperature $\beta=1/T$
will eventually be taken to be infinite for the lowest-energy resonant state in the sense of the real part of the energy.
Introducing the $S$ matrix
\begin{align}
    S(\beta)=T_{\tau}\exp\left[-\int_0^{\beta}d\tau\,V_{\rm ex}(\tau)\right],
\end{align}
we have
\begin{align}
    Z={\rm Tr}\left[e^{-\beta(H_{>}+H_{<})}S(\beta)\right],
\end{align}
where
\begin{align}
    V_{\rm ex}(\tau)=e^{\tau(H_{>}+H_{<})}V_{\rm ex} e^{-\tau(H_{>}+H_{<})}.
\end{align}
The partial trace of the slow mode gives
\begin{align}
    Z ={\rm Tr}_{<}\left[e^{-\beta (H_{<}+V_{<}^{\rm eff})}\right],
\end{align}
where $V_{<}^{\rm eff}$ is the effective interaction in the slow modes and ${\rm Tr}_{<(>)}[\cdots]$ is the partial trace for the slow (fast) mode.
Then, we find
\begin{align}
   e^{-\beta V_{<}^{\rm eff}}= {\rm Tr}_{>}\left[e^{-\beta H_{>}}S(\beta)\right]
\end{align}
The perturbative expansion of $S(\beta)$ allows us to evaluate the interaction correction in each RG flow step.
The effective interaction is given by the second-order perturbation process 
\begin{align}
\label{eq:b7}
    V_{<}^{\rm eff}=-\frac{1}{2\beta}\int_0^{\beta}d\tau_1\int_0^{\beta}d\tau_2
    \langle T_\tau [V_{\rm ex}(\tau_1)V_{\rm ex}(\tau_2)]\rangle_{>},
\end{align}
where $\langle\cdots\rangle_{>}={\rm Tr}_{>}\left[e^{-\beta H_{>}}\cdots\right]/{\rm Tr}_{>}\left[e^{-\beta H_{>}}\right]$.
We consider the contact-type interaction which is given in the momentum space by
\begin{align}
    V_{\rm c}=g_\Lambda\sum_{\bm{k},\bm{k}'}
    c_{\bm{k}\up}^\dag
    c_{-\bm{k}\down}^\dag
    c_{-\bm{k}'\down}
    c_{-\bm{k}'\up},
\end{align}
where 
$c_{\bm{k}\sigma}$ is the annihilation operator.
Using the decomposition $c_{\bm{k}\sigma}=c_{\bm{k}\sigma>}+c_{\bm{k}\sigma<}$ with the operators of each mode, we separate $V_{\rm c}$ in the form of
\begin{align}
    V_{\rm c}=V_{<}+V_{>}+V_{\rm ex}.
\end{align}
Note that $V_{<}$ is the interaction in the slow mode before the renormalization.
We have the exchange term between the fast and slow modes
\begin{widetext}
    \begin{align}
    V_{\rm ex}&=g_\Lambda\sum_{|\bm{k}|\leq \Lambda-d\Lambda}
    \sum_{\Lambda-d\Lambda\leq|\bm{k}'|\leq\Lambda}
    c_{\bm{k}\up<}^\dag
    c_{-\bm{k}\down<}^\dag
    c_{-\bm{k}'\down>}
    c_{-\bm{k}'\up>}%\cr
    %&
    +g_\Lambda
    \sum_{\Lambda-d\Lambda\leq|\bm{k}|\leq\Lambda}
    \sum_{|\bm{k}'|\leq \Lambda-d\Lambda}
    c_{\bm{k}\up>}^\dag
    c_{-\bm{k}\down>}^\dag
    c_{-\bm{k}'\down<}
    c_{-\bm{k}'\up<}.
\end{align}
\end{widetext}
The second-order perturbation given by Eq.~\eqref{eq:b7} leads to 
\begin{align}
\label{eq:b12}
    g_{\Lambda-d\Lambda}=g_{\Lambda}-g_{\Lambda}^2\sum_{\Lambda-d\Lambda<|\bm{p}|<\Lambda}\frac{2\mu}{p^2}+O(d\Lambda^2),
\end{align}
where the left-hand side is the updated coupling in the rescaling flow.
Namely, the running coupling is updated on the basis of the information where the inelastic two-body loss with strength of $g_{i,\Lambda}$ does not occur during the rescaling process.
We thus obtain Eq.~\eqref{eq:19} in the main text.

\section{Two-body resonance in $d=1$ and $d=3$}
\label{app:d}
In this appendix, we present the two-body resonance energy in $d=1$ and $d=3$.
The measurement-induced resonant state with a pure imaginary coupling is also discussed.
The two-body resonance energy in $d=1$ is obtained from the pole of the $T$ matrix given by
\begin{align}
  1-g_{\Lambda_0}\Pi(E)=  1+g_{\Lambda_0}\frac{\mu}{\pi\sqrt{2\mu E}}=0,
\end{align}
where the limit $\Lambda\rightarrow \infty$ is taken in Eq.~\eqref{eq:18}.
We thus find the resonance energy $E\equiv E_{\rm R}-i\Gamma$ as
\begin{align}
    E_{\rm R}-i\Gamma&=-\frac{\mu g_{\Lambda_0}^2}{2}\cr
    &=-\frac{\mu(g_{r,\Lambda_0}^2-g_{i,\Lambda_0}^2)}{2}-i\mu g_{r,\Lambda_0}g_{i,\Lambda_0}.
\end{align}
In terms of $U_0=\frac{2\mu g_{\Lambda_0}}{\pi\Lambda_0}$ for $d=1$,
we obtain
\begin{align}
    \frac{2\mu(E_{\rm R}-i\Gamma)}{\Lambda_0^2}=-\frac{\mu^2g_{\Lambda_0}^2}{\Lambda^2}
    =-\frac{\pi^2}{4}U_0^2.
\end{align}
For $U_0=-iU_{i,0}$ in $d=1$, we have
\begin{align}
\label{eq:d4_2}
    \frac{2\mu E_{\rm R}}{\Lambda_0^2}=\frac{\pi^2}{4}U_{i,0}^2,
    \quad \frac{2\mu \Gamma}{\Lambda_0^2}=0.
\end{align}
Since Eq.~\eqref{eq:d4_2} gives $E_{\rm R}>0$ and $\Gamma=0$,
there is no localized resonant state in $d=1$ that is induced solely by the imaginary coupling.

For $d=3$, the pole of the $T$ matrix reads
\begin{align}
    1+g\left(\frac{\mu \Lambda_0}{\pi^2}+i\frac{\mu}{2\pi}\sqrt{2\mu E}\right)=0.
\end{align}
Then, the resonance energy is given by
\begin{align}
 E_{\rm R}-i\Gamma
   % &=-\frac{1}{m}\left[\frac{4\pi}{m|g_{\Lambda_0}|^2}(g_{r,\Lambda_0}+ig_{i,\Lambda_0})+\frac{2\Lambda_0}{\pi}\right]^2\cr
    &=-\frac{1}{2\mu}
    \left(\frac{2\pi g_{r,\Lambda_0}}{\mu|g_{\Lambda_0}|^2}+\frac{2\Lambda_0}{\pi}\right)^2+\frac{2\pi^2g_{i,\Lambda_0}^2}{\mu^3|g_{\Lambda_0}|^4}\cr
    &\quad-i\frac{2\pi g_{i,\Lambda_0}}{\mu^2|g_{\Lambda_0}|^2}\left(\frac{2\pi g_{r,\Lambda_0}}{\mu|g_{\Lambda_0}|^2}+\frac{2\Lambda_0}{\pi}\right).
\end{align}
Using $U_0=\frac{1}{\pi^2}\mu g_{\Lambda_0}\Lambda_0$ for $d=3$,
we obtain
\begin{align}
    \frac{2\mu (E_{\rm R}-i\Gamma)}{\Lambda_0^2}
    &=-\left(\frac{2\pi}{\mu g_{\Lambda_0}\Lambda_0}+\frac{2}{\pi}\right)^2\cr
    &=-\frac{4}{\pi^2}\left(\frac{1}{U_0}+1\right)^2.
\end{align}
For $U_{0}=-iU_{i,0}$ in $d=3$, we obtain
\begin{align}
    \frac{2\mu E_{\rm R}}{\Lambda_0^2}=-\frac{4}{\pi^2}\left(1-\frac{1}{U_{i,0}^2}\right),\quad
    \frac{2\mu \Gamma}{\Lambda_0^2}=\frac{8}{\pi^2}\frac{1}{U_{i,0}}.
\end{align}
The sign of $E_{\rm R}$ changes at $U_{i,0}=1$ and $\Gamma$ is always nonzero for arbitrary finite $U_{i,0}$.

\section{Localized two-body wave function in $d=2$}
\label{app:e}
In this appendix, we show the form of the resonant wave function.
To this end,
we start from the non-Hermitian Schr\"{o}dinger equation:
\begin{align}
\label{eq:d1}
    \left[-\frac{\nabla^2}{2\mu}+g_{\Lambda_0}\delta(\bm{r})\right]\Psi(\bm{r})=E\Psi(\bm{r}),
\end{align}
where $\Psi(\bm{r})$ is the two-body wave function with relative coordinate $\bm{r}$.
We expand $\Psi(\bm{r})$ with respect to the plane wave as
\begin{align}
    \Psi(\bm{r})=
    \sum_{\bm{k}}
    C_{\bm{k}}
    e^{i\bm{k}\cdot\bm{r}},
\end{align}
where $C_{\bm{k}}$ is the Fourier coefficient of $\Psi(\bm{r})$.
Then Eq.~\eqref{eq:d1} can be rewritten as
\begin{align}
    \frac{1}{\sqrt{V}}\sum_{\bm{k}}\left[\frac{k^2}{2\mu}+g_{\Lambda_0}\delta(\bm{r})-E\right]
    C_{\bm{k}}
    e^{i\bm{k}\cdot\bm{r}}
    =0.
\end{align}
Multiplying $e^{-i\bm{p}\cdot\bm{r}}$ and integrating them with respect to $\bm{r}$, we obtain the self-consistent equation
\begin{align}
\label{eq:d4}
    \left(\frac{p^2}{2\mu}-E\right)C_{\bm{p}}=-g_{\Lambda_0}\sum_{\bm{k}}C_{\bm{k}}.
\end{align}
We thus obtain
\begin{align}
    C_{\bm{p}}=-\frac{g_{\Lambda_0}\Psi(\bm{0})}{p^2/2\mu-E}.
\end{align}
Note that $\Psi(\bm{0})$ actually involves ultraviolet divergence, and hence we introduce the cutoff regularization with $\Lambda_0$. 
Combining them, we have
\begin{align}
\Psi(\bm{r})=-g_{\Lambda_0}\sum_{\bm{k}}\frac{\Psi(\bm{0})}{k^2/2\mu-E}e^{i\bm{k}\cdot\bm{r}}.
\end{align}
The non-trivial solution with $\Psi(\bm{0})\neq 0$ is obtained from
\begin{align}
    1=-g_{\Lambda_0}\sum_{\bm{k}}\frac{1}{k^2/2\mu-E}\equiv g_{\Lambda_0}\Pi(E),
\end{align}
which is consistent with the pole condition of the $T$ matrix.

In $d=2$,
the real-space two-body wave function reads
\begin{align}
    \Psi(\bm{r})
    =\frac{\mu}{\pi}\Psi(\bm{0})K_0(r/\xi),
\end{align}
where $K_0(r/\xi)$ 
is the modified Bessel function of the second kind and $\xi:=1/\sqrt{-2\mu E}$.
Thus, the long-distance asymptotic behavior of the two-body wave function is given by
$\Psi(\bm{r})\sim e^{-r/\xi}/\sqrt{r}$.

\bibliography{reference}

\end{document}